\documentclass[fleqn,usenatbib]{mnras}

\usepackage{mathptmx}
\usepackage{graphicx}
\usepackage{color}
\usepackage{amssymb}
\usepackage{amsmath}
\usepackage{xspace}
\usepackage{balance}
\usepackage[T1]{fontenc}

\DeclareRobustCommand{\VAN}[3]{#2}
\let\VANthebibliography\thebibliography
\def\thebibliography{\DeclareRobustCommand{\VAN}[3]{##3}\VANthebibliography}

\newcommand{\yr}    	{\ifmmode \mathrm{yr} \else yr\fi}
\newcommand{\mpc}   	{\ifmmode \,\mathrm{Mpc}^{-3} \else \,Mpc$^{-3}$\fi}
\newcommand{\Msun}	    {\ifmmode \,\mathrm M_{\odot} \else $\,\mathrm M_{\odot}$\fi\xspace}
\newcommand{\Zsun}	    {\ifmmode \,\mathrm Z_{\odot} \else $\,\mathrm Z_{\odot}$\fi\xspace}
\newcommand{\Mhalo} 	{\ifmmode M_{\mathrm{halo}} \else $M_{\mathrm{halo}}$\fi\xspace}
\newcommand{\Rvir}  	{\ifmmode R_{200} \else $R_{200}$\fi\xspace}
\newcommand{\Mstar}	    {\ifmmode {M}_{\star} \else ${M}_{\star}$\fi\xspace}
\newcommand{\Mvir}	    {\ifmmode M_{\mathrm{halo}} \else $M_{\rm halo}$ \fi\xspace}
\newcommand{\Htwo}  	{\ifmmode {\rm H}_{2} \else ${\rm H}_{2}$ \fi}
\newcommand{\nH}    	{\ifmmode {n}_{\rm H} \else ${n}_{\rm H}$ \fi}
\newcommand{\cc}	    {\ifmmode {\rm cm}^{-3} \else ${\rm cm}^{-3}$ \fi}
\newcommand{\mperyr}	{\ifmmode \Msun{\rm yr}^{-1} \else $\Msun{\rm yr}^{-1}$ \fi}
\newcommand{\arepo}	    {{\small AREPO}\xspace}
\newcommand{\lyra}  	{{\small LYRA}\xspace}

\newcommand{\tkms}      {$+3\,\mathrm{km\,s}^{-1}$}

\title[LYRA project II]{LYRA II: Cosmological dwarf galaxy formation with inhomogeneous Population III enrichment}
\author[T. A. Gutcke et al.]{Thales
  A. Gutcke$^1$\thanks{thales@mpa-garching.mpg.de},
  R\"{u}diger Pakmor$^1$, Thorsten Naab$^1$,
  Volker Springel$^1$\\
$^{1}$Max-Planck-Institut f\"ur Astrophysik, Karl-Schwarzschild-Str. 1, D-85748, Garching, Germany
}

\begin{document}
\pagerange{\pageref{firstpage}--\pageref{lastpage}} \pubyear{---}
\maketitle
\label{firstpage}
%==========================================================================%
\begin{abstract}
We present the simulation of a $2\times10^{9}\Msun$ halo mass cosmological dwarf galaxy run to $z=0$ at $4$ solar mass gas resolution with resolved supernova feedback. We compare three simple subgrid implementations for the inhomogeneous chemical enrichment from Population III stars and compare them to constraints from Local Group dwarf galaxies. The employed model, \lyra, is a novel high resolution galaxy formation model built for the moving mesh code \arepo, which is marked by a resolved multi-phase interstellar medium, single stars and individual supernova events. The resulting reionization relic is characterized by a short ($<1.5$\,Gyr) star formation history that is repeatedly brought to a standstill by violent bursts of feedback. Star formation is reignited for a short duration due to a merger at $z\approx4$ and then again at $z\approx0.2-0$ after sustained gas accretion. Our model $z=0$ galaxy matches the stellar mass, size, stellar kinematics and metallicity relations of Local Group dwarf galaxies well. The dark matter profile does not exhibit a core in any version of the model. We show that the host halo masses of Population III stars affect the assembly history of dwarf galaxies. This manifests itself through the initial gaseous collapse in the progenitor halos, affecting the central density of the stellar component and through the accretion of luminous substructure.
\end{abstract}
%==========================================================================%
\section{Introduction}
%
%==========================================================================%

Dwarf galaxies are increasingly becoming a subject of study both for constraining dark matter and galaxy formation. Their small sizes and low mass provide an ideal environment to probe if and to what extent dark matter profiles can be affected by baryonic physics. Their shallow potentials are extremely sensitive to feedback processes such as supernovae, so they can be treated as record keepers for processes long passed. Also, their chemical composition is a great test bed for early star formation (SF). The minute star formation rates of dwarf galaxies create only little metals, so individual enrichment events may be distinguished in the stellar or gas metallicity records.

The interest in dwarf galaxies truly began after the Sloan Digital Sky Survey discovered a large population of ultra-faint (UFD) and classical dwarfs in the Local Group \citep[LG, e.g.][]{Willman2005, Belokurov2006}, see \cite{Simon2019} for a review. Thus, the past decade has given rise to many studies achieving substantial progress in our theoretical understanding of dwarf galaxies and their place in galaxy evolution \citep[e.g.][]{Governato2010, Governato2015, Sawala2011, Munshi2013, Simpson2013, Trujillo-Gomez2015, Onorbe2015, Fitts2017, Maccio2017, Frings2017, Anora2021}. The ancient stellar populations with either no or very little ongoing SF at the present day observed in most LG dwarfs can be explained by a combination of gas heating by reionization and subsequent supernovae heating and ejection of any remaining self-shielded gas.     

Connecting early SF with LG observations, \cite{Jeon2017} present {\small GADGET} simulations of six UFDs in the halo mass range $1-4 \times 10^9$\,\Msun. Their simulations include a prescription for Population III (PopIII) stars with an altered IMF. {{They divide the stars into ``in-situ'' and ``ex-situ'' groups, depending on whether the stars were born within or outside of the virial radius of the main halo at any time, respectively. They proceed to}} show that the ex-situ fraction is much higher in the lowest mass halos than in more massive systems. They also point out that some halos can form Population II (PopII) stars in their first cycle of SF due to metal contamination from SNe of neighboring halos (since halos are very close together at high redshift).

Using the FIRE-2 galaxy formation model, \cite{Wheeler2019} present three UFD analogs. The authors show that many $z=0$ constraints from observations are well reproduced in their simulated ultra-faint dwarfs. One notable outlier to this is the mass-metallicity relation, where the simulations under-predict the metals by more than an order of magnitude. They suggest that this may be ameliorated by following PopIII enrichment and the individual evolutionary pathway of stars rather than attempting to resolve the collective effects further.

Recently, it has become computationally feasible to run high resolution interstellar medium (ISM) models within cosmological zoom-in simulations of field dwarfs.
The EDGE project \citep{Agertz2020} presents a $10^9$\Msun dwarf galaxy for which the authors resolve the multi-phase ISM, individual supernovae (SNe), and radiation via radiation transfer \citep[with RAMSES-RT,][]{Rosdahl2013} from young stars in a cosmological context. They conclude that the galactic averaged stellar metallicity is a sensitive indicator able to discriminate between feedback models. Subsequent analysis of additional EDGE simulations is presented in \cite{Orkney2021, Rey2019, Rey2020}.

While the increase in resolution and the addition of more detailed gas physics with resolved supernovae enables a deeper insight into the ISM and gaseous outflows, it also poses a new set of questions to cosmological hydrodynamical simulations \citep[see][]{Naab2017}. These high resolution simulations become more sensitive to local enrichment events and the timing of reionization. This is especially true for low mass dwarf galaxies.
%An interesting avenue of investigation is to connect the high redshift early enrichment of small halos with the $z=0$ characteristics of LG dwarf galaxies. 
Their SF rates are extremely low ($10^{-6} - 10^{-2} \Msun\mathrm{yr}^{-1}$), and quenching through reionization is expected to set in within the first gigayear after formation. Thus, the stellar and chemical properties of such systems should be highly sensitive to the timing of the onset of SF and the characteristics of the first halos that host the first stars, so-called PopIII stars \citep{Bromm2004}. \cite{Machacek2001} showed that a constant Lyman-Werner background radiation can suppress PopIII SF in small halos and thus delay the first stars until more massive halos form. The authors give a fitting function to their simulation results that predicts a redshift-dependent host halo mass for the first stars in the range $1-3\times10^6\Msun$.

A more detailed PopIII model is used in \cite{Wise2012}, however only evolved to $z=7$. They compute simulations that distinguish PopII and PopIII star formation. From their simulations, they show that the expected average halo metallicity from PopIII SNe is strongly dependent on the halo mass. They also demonstrate that {{their halos begin by forming PopIII stars that expel the first metals. These first metals initiate the transition to PopII star formation around a halo mass of $\gtrsim10^7\Msun$}}. However, \cite{Skinner2020} show that $\mathrm{H}_2$-shielding can significantly alter the amount of radiation that penetrates these halos, thus decreasing the mass requirements. They predict a redshift-dependent mean host halo mass of PopIII stars that ranges from $3\times10^5$ to $10^7\Msun$. The dependency is driven by the self-consistently simulated radiation background that increases over time as more stars form and emit. Additionally, \cite{Schauer2019} show that including streaming velocities can alter the expected mass of the first star forming halos. \cite{Schauer2021} proceed to postulate SF outside of the first halos with the aid of metals that escape the virial radius and provide additional cooling avenues in the inter-galactic medium.
 
Our work aims to {{investigate}} a simple PopIII enrichment model {{and compare its results}} with observations of $z=0$ LG dwarfs. To this end we examine the evolution and stellar properties of a simulated dwarf at very high resolution. Since $10^9\Msun$ dwarfs are expected to be hierarchically composed of smaller halos, we attempt to show how the merging of different stellar mass progenitors can build different SFHs and metallicity distributions.

We compare our simulations with dwarf galaxies observed as satellites to the Milky Way and as satellites to Andromeda. For these we use the dataset updated in 2021 originally provided by \cite{McConnachie2012}.  Since our dwarf is chosen to be an isolated system, the Solitary Local (Solo) Dwarf Galaxy Survey, a sample of isolated dwarf galaxies within $3$~Mpc presented in \cite{Higgs2016} provides the closest observational counterparts.
%
%This includes compiled data from ...
%
 Using the Solo survey, \cite{HiggsII2021} take a closer look at the stellar structure and substructure of this isolated sample. They do not detect any globular clusters or other luminous substructure within their field of view (about half a degree to each side of each object). \cite{Dooley2017} calculate the probability for Local Group field dwarfs to host stellar substructure. They use the Caterpillar dark matter-only simulations of Milky Way analogs and their surroundings with various abundance matching models. Depending on the mass, these dwarfs are expected to host 0 to 3 satellites. \cite{HiggsIV2021} further compare the isolated sample with the satellite samples from the Milky Way and Andromeda. 

An interesting aspect of certain observed dwarf galaxies is ongoing SF at the present day. The literature provides various simulations of dwarf galaxies that are initially quenched through reionization and later reignite SF. Yet the mechanism held responsible for the late-time SF in dwarfs varies between authors. \cite{Wright2019} present simulations in which SF is triggered after the ISM is compressed due to interactions with shock waves or gas streams from nearby galaxies. \cite{Rey2020} on the other hand show late-time SF caused by a slow accretion of gas following reionization that is finally able to rejuvenate the galaxy 6-8 billion years later. In the following, we will confirm the slow accretion scenario and also show merger-driven re-ignition of SF post-reionization.

In this work, we analyze dwarf galaxy simulations run with the \lyra model at $4$ solar mass resolution to $z=0$. The aim is to understand the formation and evolution of dwarf galaxies with as few assumptions as possible, and thus to learn about the complex interplay of physics that leads to their current properties. We -- for the first time -- are able to follow the assembly and evolution from the earliest enrichment by PopIII stars to the $z=0$ properties of the smallest known galaxies in the Universe by simulating the multi-phase ISM and individually resolved supernova blastwaves of all progenitors and their subsequent merging. We would like to emphasize the fact that cosmological simulations in this extremely high resolution regime have only recently become feasible and are still rare. This new level of precision may be viewed as an important step towards a more quantitative understanding of feedback. Also, in this regime small-scale processes that have mostly been disregarded by previous cosmological simulations begin to have significant impact and require accurate modelling (be it sub-grid or otherwise). This makes the comparison of different simulation efforts at this scale particularly vital.

The paper is organized as follows. In Sec.~\ref{sec:model} we present the employed \lyra model and its new modifications. In Sec.~\ref{sec:haloprops} we first present the general halo properties of the simulations. Then we describe the star formation and subsequent quenching in Sec.~\ref{sec:sf}. Next, we verify our model by comparing the $z=0$ properties with LG dwarf galaxy data in Sec.~\ref{sec:z0}. Lastly, we present two main predictions of the model. In Sec.~\ref{sec:dmprofile} we show that the dark matter profile does not develop a core. Finally, in Sec.~\ref{sec:popiii} we discuss the effect that PopIII enrichment has on the luminous substructures of the dwarf. In Sec.~\ref{sec:conclusion}, we summarize our main conclusions and discuss differences to other simulation work.

During the following, we use the $\Lambda$CDM cosmological parameters $\Omega_{\Lambda} = 0.693$, $\Omega_0 = 0.307$, $\Omega_b = 0.048$ and $h=0.6777$ \citep{PlanckCollaboration2014}. We take $Z_\odot = 0.01337$, following \cite{Asplund2009}.
%
%==========================================================================%
\section{The LYRA model}
\label{sec:model}
%==========================================================================%
\begin{figure}
  \includegraphics[width=\linewidth]{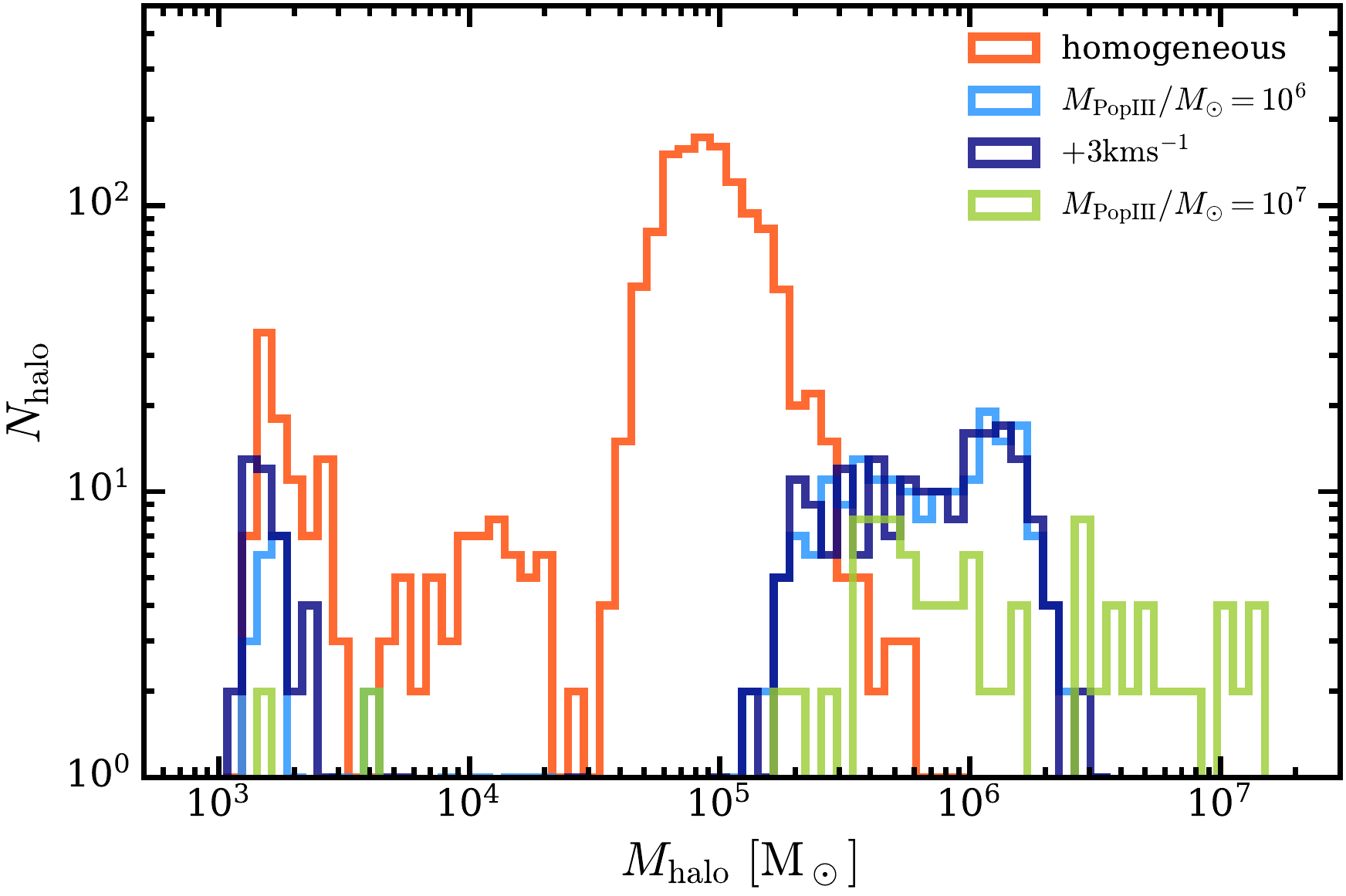}
  \caption[]{Distribution function of halo masses at the time when they host their first star. The homogeneous model mainly begins forming stars in $10^5\Msun$ halos. These then enrich smaller halos in their surroundings, which leads to the tail to lower masses. The other two simulations have tails to higher masses than their $M_\mathrm{PopIII}$ value. This is due to the discrete time intervals in which we run the on-the-fly halo finder, which allows some halos to grow larger than the threshold.}
  \label{fig:firstsubs}
\end{figure}

For the study of dwarf galaxies, we use the \lyra model \citep{Gutcke2021}, a novel galaxy formation code built within the magneto-hydrodynamics code \arepo \citep{Springel2010,Pakmor2016,Weinberger2020}. \lyra resolves the multi-phase ISM down to 10\,K, self-consistently modelling the formation of molecular clouds. Individual stars drawn from the IMF are evolved with individual, mass-dependent lifetimes. At death, the stars return mass, metals and energy (in the case of SNe) back to the ISM. These are mixed into the pre-existing ISM structure, creating outflows and enriching the clouds for future cycles of SF. Our model is characterized by a very high gas mass resolution, adaptive timestepping and SN blastwaves for which the Sedov phase is resolved \citep[see also][for similar models]{Kim2015,Kim2017,Hu2016,Hu2017}. These in concert are able to resolve and follow the over-pressurized, supersonic outflows created by the clustered injection of SNe. %\citep[see][for a detailed analysis]{Gutcke2021}.

We summarize the most relevant aspects of the model here, but refer the reader to \cite{Gutcke2021} for all other details. In its cosmological mode, \arepo integrates the equations of motion in a Friedman-Robertson-Walker metric for which we use the latest cosmological parameters published by the \cite{PlanckCollaboration2014}. 
We include heating from the UV background \citep{Faucher-Giguere2020} and self-shielding according to \cite{Rahmati2013}. Gas cooling is achieved with a primordial, atomic network \citep{Katz1996} combined with interpolation of the ``UVB dust1 CR0 G0 shield1" cooling table from \cite{Ploeckinger2020}. The latter includes metal line cooling, low temperature cooling, molecular cooling (with H$_2$), a dust model, and accounts for self-shielding in the low temperature regime. The corresponding table has four dimensions: redshift, density, temperature and metallicity.

Stars are formed following a Schmidt-relation with a density threshold of $\nH>10^3\,\cc$, and an efficiency parameter of $0.02$. We force star formation when the density rises above $\nH>10^4\,\cc$. Stellar masses are drawn from a Kroupa IMF \citep{Kroupa2001}. The IMF is resolved down to $M_{\star\mathrm{, min}}=4\Msun$. Below this value, the IMF is considered \textit{unresolved}, and the corresponding integrated mass is distributed into stars with mass equal to $M_{\star\mathrm{, min}}$. {{Stars below $M_{\star\mathrm{, min}}$ do not return mass or metals. Since this is a very minor contribution, we choose to neglect it in favor of computational speed. Instead, the presence of these stars has a purely dynamical effect on the simulation.}}

Stars above $M_{\star\mathrm{, min}}$ have characteristic evolutionary pathways according to their initial formation mass and formation metallicity. These set their lifetime \citep{Portinari1998} and determine one of three pathways for stellar death: Asymptotic giant branch (AGB), SNe, or direct collapse black holes (BHs). AGB and direct collapse black holes return mass and metals, while SNe additionally return energy to the ISM. The adopted SNe and direct collapse BH returns follow the SN explosion models presented in \cite{Sukhbold2016}. The SNe energy is therefore variable and depends on the initial stellar mass. AGB returns follow \cite{Karakas2018}.

% \begin{figure*}
%   \includegraphics[width=\linewidth]{images/SN_AGB_densities.pdf}
%   \caption[]{Density distribution of AGB, SN and direct collapse BHs. The distributions are fully consistent with tests shown in \citet{Gutcke2021}}
%   \label{fig:SNdensity}
% \end{figure*}

\subsection{Population III enrichment}
PopIII stars have never been observed directly, but are postulated to be the first stars that form in the Universe from quasi metal-free gas \citep{Bromm2013}. The major coolant is $\mathrm{H}_2$. They are expected to have masses of around $100\,\Msun$ \citep{Abel2002} and become energetic SN with high chemical yields. Thus, their existence is vital for all subsequent galaxy and star formation, because the metals produced by them allow gas to cool through many more channels. Most cosmological models that run simulations to $z=0$ do not, however, include PopIII stars. Instead of modelling PopIII stars directly, most models set an initial metallicity floor (often set to [Fe/H]~$= -4$) throughout the gas in the simulation to mimic the results of PopIII chemical pre-processing. 

However, this is an approximation that does not take the inhomogeneity of the initial enrichment into account. While PopIII stars still have to be discovered observationally, there are various attempts at theoretically constraining the mass of the host halos in which these stars should have formed \citep[e.g][]{Machacek2001, Wise2012, Jeon2017, Skinner2020}. Depending on the assumptions made in each model, the predicted halo masses lie between $10^5$ to $10^7~\Msun$, with a possible redshift dependence. The expected metallicity of PopIII-hosting halos after their star has died is between $\log Z_{\rm PopIII} \approx-5$ and $-4$.

We have built a simple subgrid model for the initial enrichment of the gas to mimic PopIII enrichment. To this end, the beginning state of all gas in the simulation is set to metal-free. Then, only gas in halos that have passed a certain mass threshold, $M_{\rm PopIII}$, are enriched to $Z_{\rm PopIII}$. For definiteness, we adopt $\log Z_{\rm PopIII}=-4$. To test the effect of different values of $M_{\rm PopIII}$, we run the same initial conditions with two values for this threshold, $10^6\,\Msun$ and $10^7\,\Msun$. Additionally, we carry out a simulation with homogeneous enrichment where all gas gets initialized to $Z_{\rm PopIII}$ from the start. A caveat of this model is that the feedback energy injected by the exploding PopIII stars is neglected. While this may have a quenching effect on SF by expelling gas, we argue that the additional cooling generated by the added metals is the larger of the two effects. 

In Fig.~\ref{fig:firstsubs}, we show the mass distribution of halos that host their first star. It is interesting to note that all three PopIII models display the majority of their halos with masses well below $M_\mathrm{PopIII}$.  While these subhalos did not pass the mass threshold given by the PopIII enrichment model, they were nonetheless able to form stars at high redshift. This is due to outflows originating from neighboring halos that expel enriched gas well beyond their own virial radii \citep[see also][for examples of this]{Jeon2017}. This enriched gas can mix with the gas in smaller halos, allowing them to obtain the ability to cool, condense gas, and finally form stars. This process happens before reionization ($z>8$), and is brought to a standstill when the increasing radiation background penetrates these small halos. {{In the figure, we also show a fourth model (\tkms) that is identical to the $10^6\,\Msun$ except that we give each star particle a velocity kick of magnitude $3\mathrm{km/s}$ in a random orientation. This kick does not significantly affect the distribution of PopIII halo masses.}}

\subsection{Modelling the luminosity}
To obtain an estimate of the luminosity and magnitude of the stars and the galaxy as a whole, we have developed a custom routine that takes LYRA's star formation implementation into account. We use the PARSEC isochrone tables provided by \cite{Bressan2012}. We interpolate these for each star according to its initial mass, age and metallicity, providing a magnitude for each star. 

LYRA splits the stellar initial mass function (IMF) into resolved and unresolved parts (see \citealt{Gutcke2021} for details). For the resolved part, the calculation of the magnitudes is straight forward. For the unresolved part, the age and metallicity of the star particle are assumed to be a proxy for the mean age and mean metallicity of the integral of the unresolved section of the IMF. Thus, these star particles are assigned a magnitude that is representative of the integrated magnitude of the lower end of the IMF at a given age and metallicity. 
%==========================================================================%
\section{Results}
\label{sec:results}
%==========================================================================%
%%%%%%%%%%%%%%%%%%%%%%%%%%%%%%%%%%%%%%%%%%%%%%%%%%%%%%%%%%%%%%%%%%%%%%%%%%%
\begin{figure}
  \includegraphics[width=\linewidth]{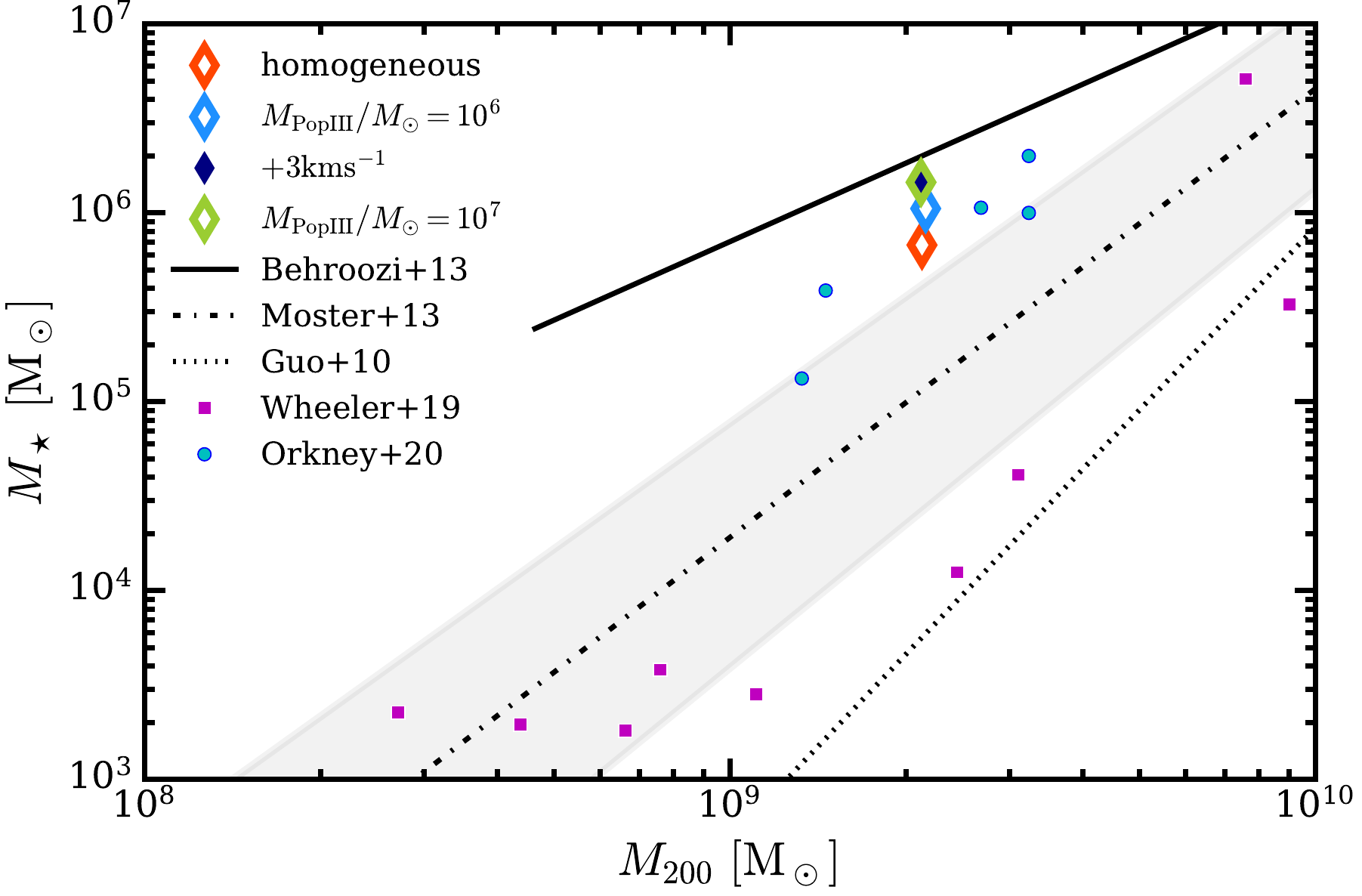}
  \includegraphics[width=\linewidth]{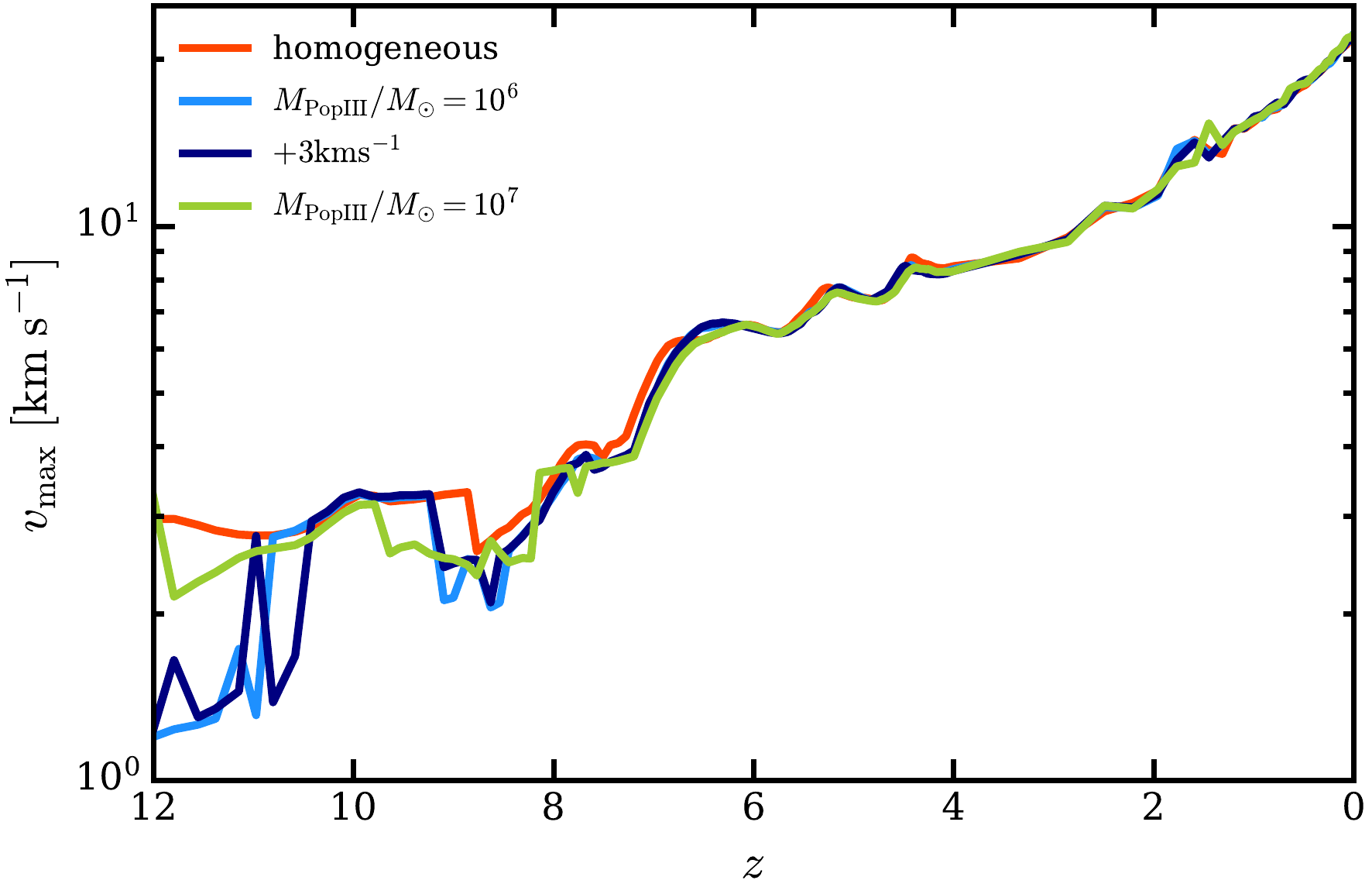}
  \caption[]{\textit{Top:} Stellar mass to halo mass relation with the $z=0$ value for our cosmological simulations. Since low mass dwarfs are mostly quenched at high redshift, they do not evolve along the abundance matching relation. Instead, they form their stars early and increase their halo mass more slowly until $z=0$. \textit{Bottom: } Evolution of the peak value of the rotation curve, $v_\mathrm{max}$, with redshift. The four different versions of our model retain the same halo properties.}
  \label{fig:SMHM}
\end{figure}
%%%%%%%%%%%%%%%%%%%%%%%%%%%%%%%%%%%%%%%%%%%%%%%%%%%%%%%%%%%%%%%%%%%%%%%%%%%
%%%%%%%%%%%%%%%%%%%%%%%%%%%%%%%%%%%%%%%%%%%%%%%%%%%%%%%%%%%%%%%%%%%%%%%%%%%
\begin{figure*}
  \includegraphics[width=\linewidth]{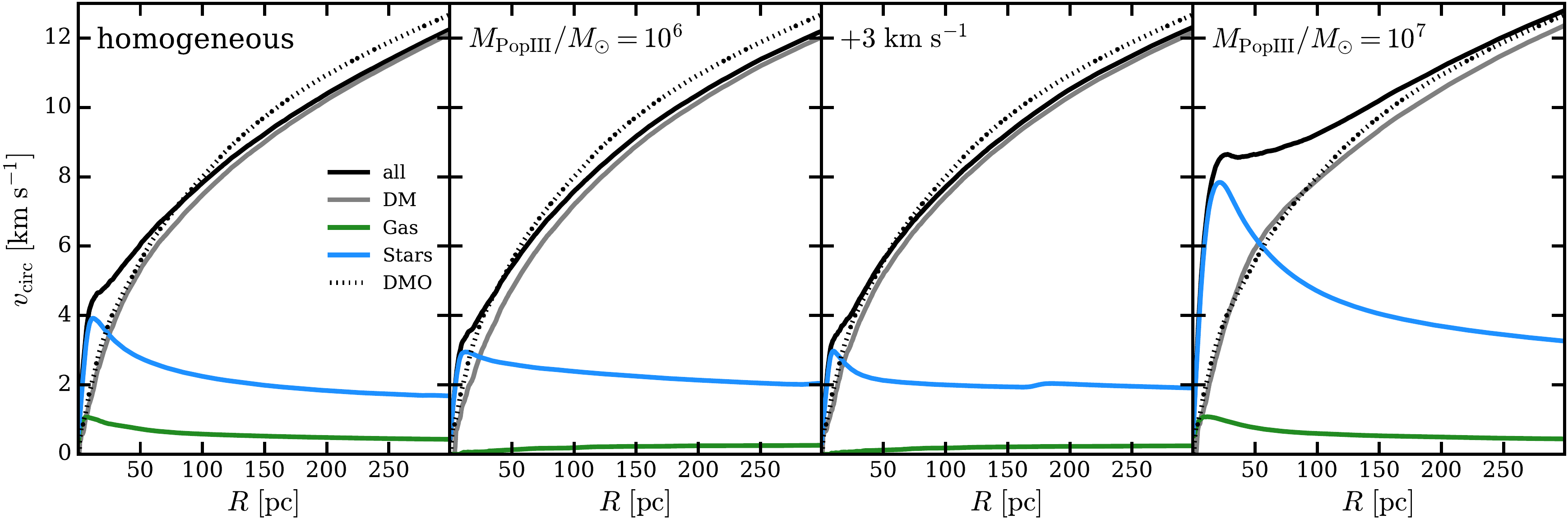}
  \caption[]{Rotation curves out to $300\,$pc at $z=0$ for each simulation. Total mass, DM, stars and gas are shown separately. The DMO curve (scaled by the cosmic baryon fraction) is also shown. It is clear that the central stellar component is quite different between the three PopIII models and that it dominates the potential out to $\sim50\,$pc. We also see that the amount of gas retained by the different models depends on the depth of the central potential, which is set by the stars.}
  \label{fig:rotcurve}
\end{figure*}
%%%%%%%%%%%%%%%%%%%%%%%%%%%%%%%%%%%%%%%%%%%%%%%%%%%%%%%%%%%%%%%%%%%%%%%%%%%
%%%%%%%%%%%%%%%%%%%%%%%%%%%%%%%%%%%%%%%%%%%%%%%%%%%%%%%%%%%%%%%%%%%%%%%%%%%
\begin{figure*}
\includegraphics[width=\linewidth]{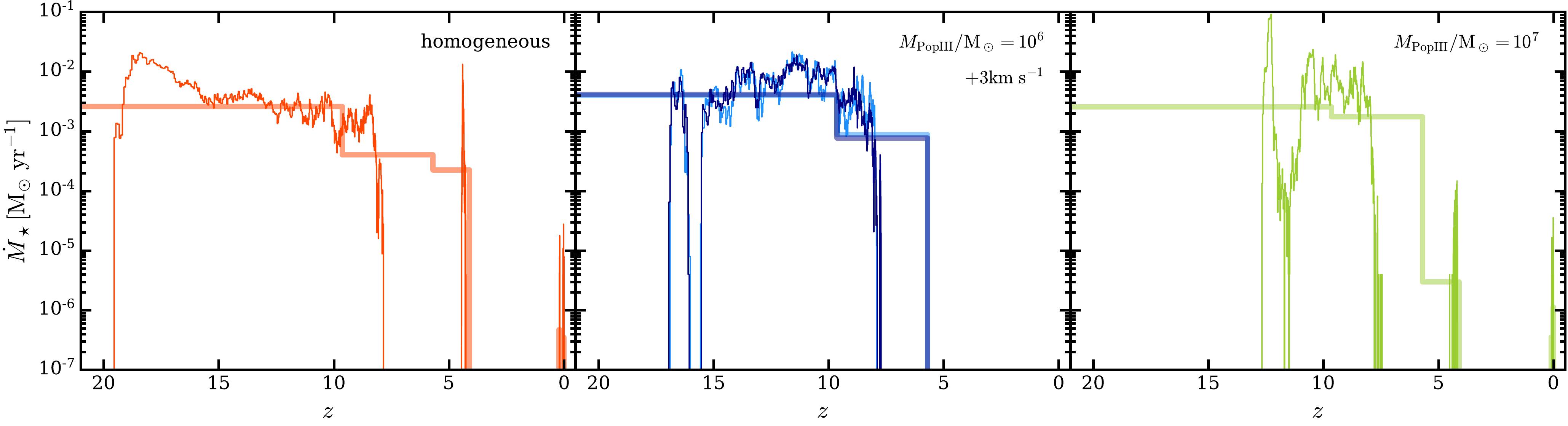}
    \caption[]{ {{Star formation history (SFH) averaged over bins of $500$\,Myr (thick solid lines) and $1$\,Myr (thin lines). Note that the $x$-axis is redshift. In the central panel, we show both ``$10^6$'' and ``\tkms''.}} The star formation histories are characterized by the quenching that occurs at the onset of reionization at around $z=8$. There is a merger at $z\approx4.5$ that reignites SF in two of the models. SF is again restarted at $z\approx0.3$ due to slow gas accretion (see Sec.~\ref{sec:dmprofile}).}
    % \caption[]{Star formation history in the first $2$\,Gyr in bins of $10$\,Myr. The histories are characterized by the quenching that occurs during reionization at around $z=8$. There is a merger at $z\approx4.5$ that reignites SF in two of the models. }
  \label{fig:sfr}
\end{figure*}
%%%%%%%%%%%%%%%%%%%%%%%%%%%%%%%%%%%%%%%%%%%%%%%%%%%%%%%%%%%%%%%%%%%%%%%%%%%
%%%%%%%%%%%%%%%%%%%%%%%%%%%%%%%%%%%%%%%%%%%%%%%%%%%%%%%%%%%%%%%%%%%%%%%%%%%
\begin{figure*}
  \includegraphics[width=0.45\textwidth]{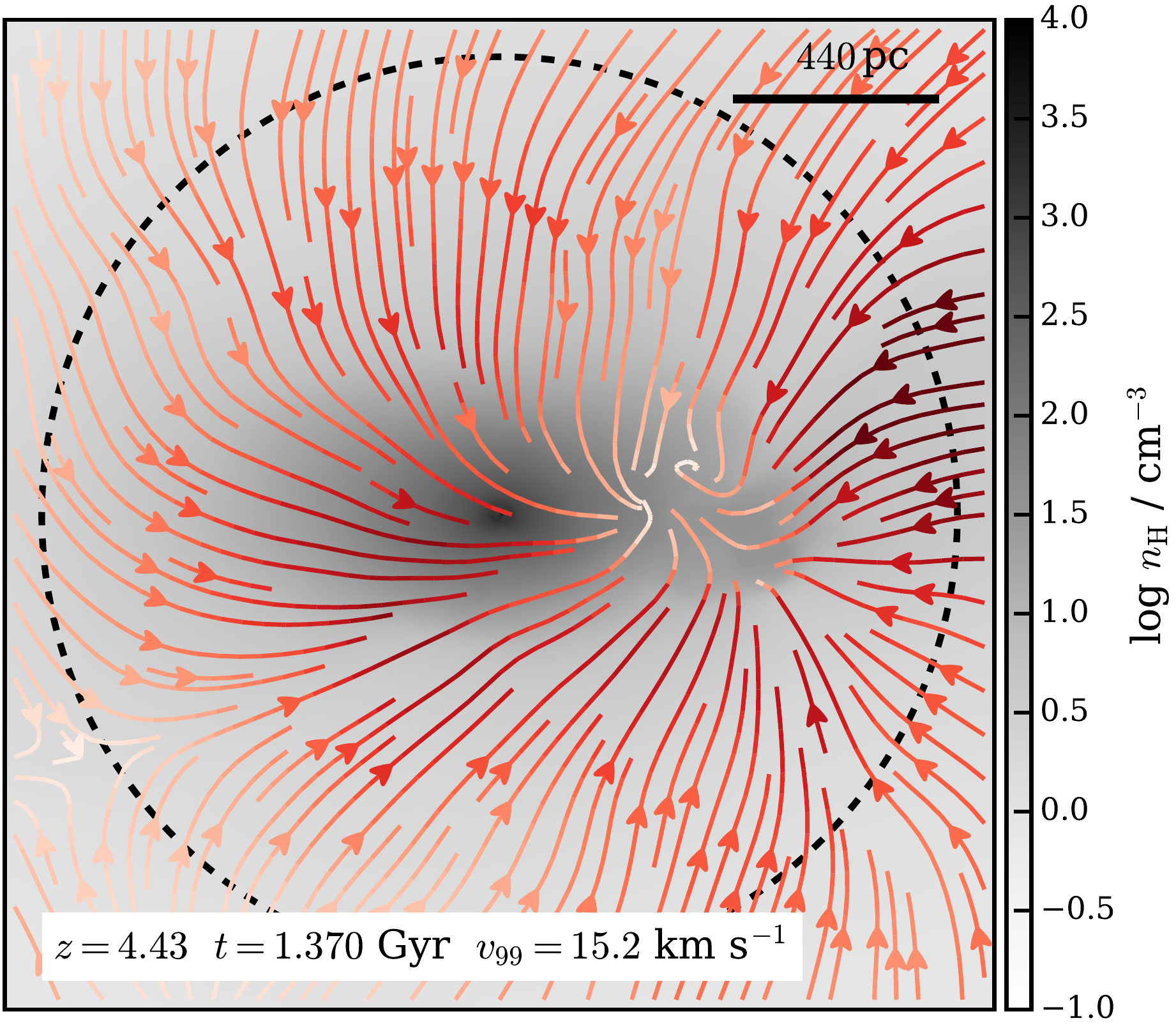}
  \includegraphics[width=0.45\textwidth]{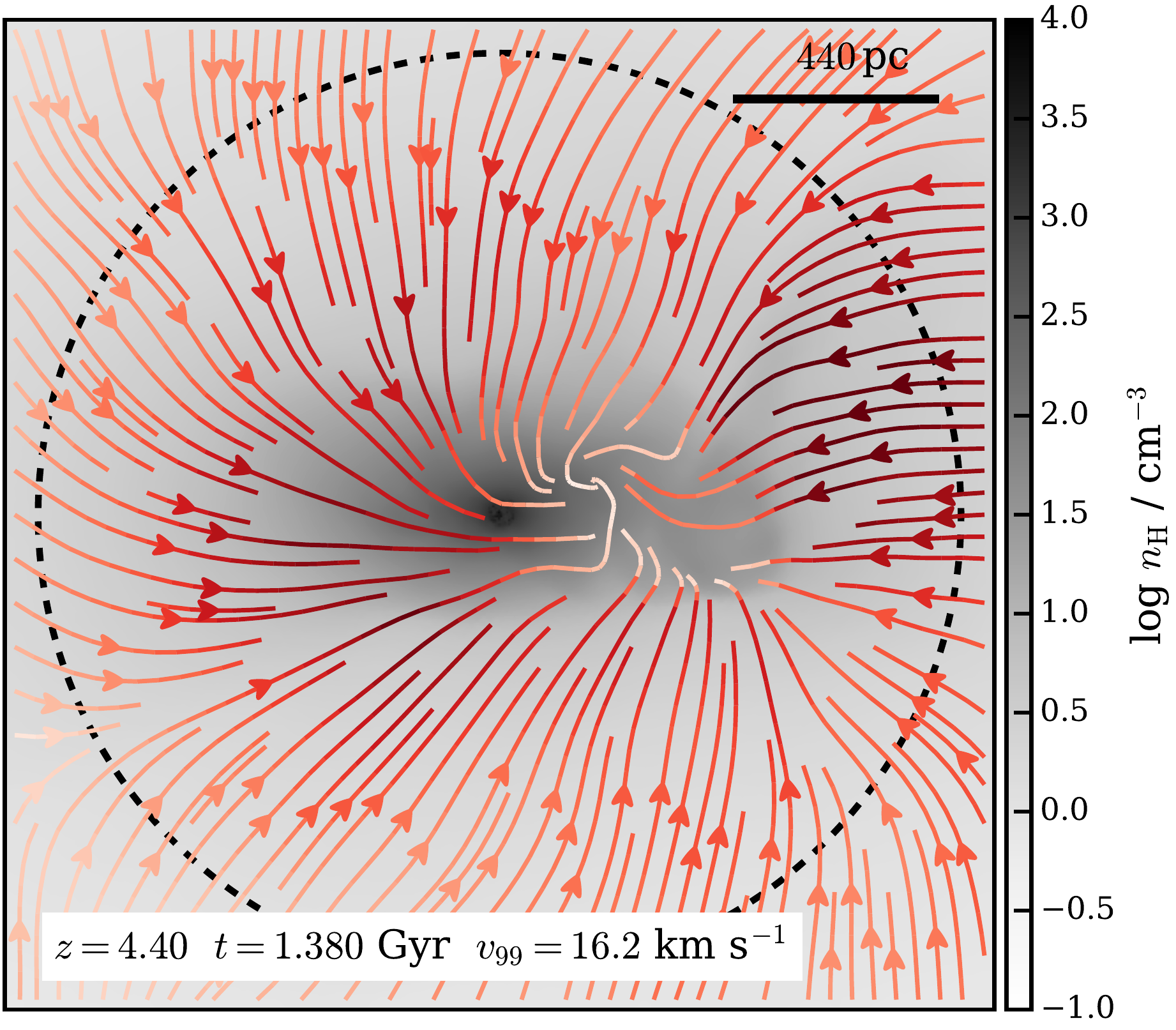}
  \includegraphics[width=0.45\textwidth]{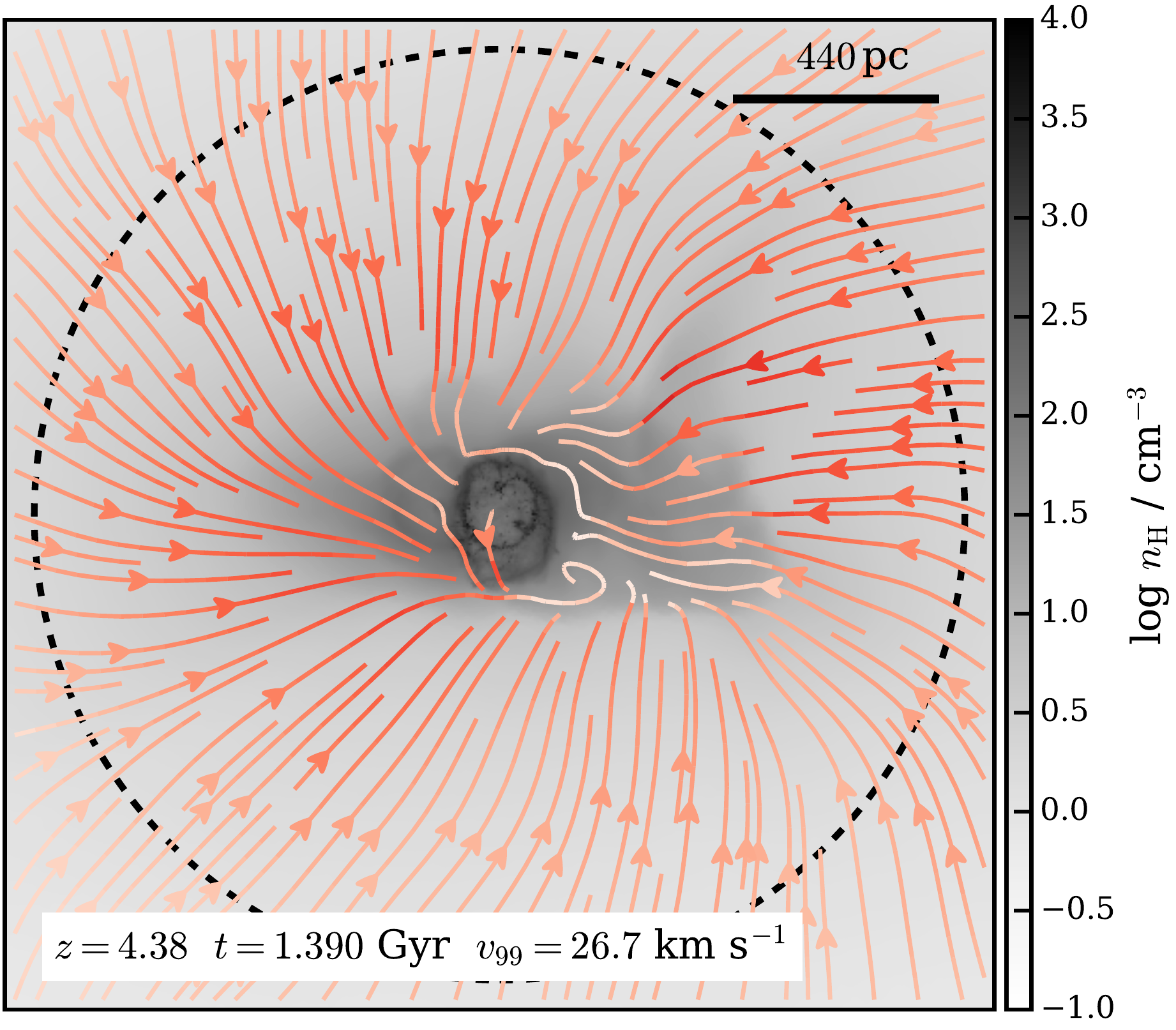}
  \includegraphics[width=0.45\textwidth]{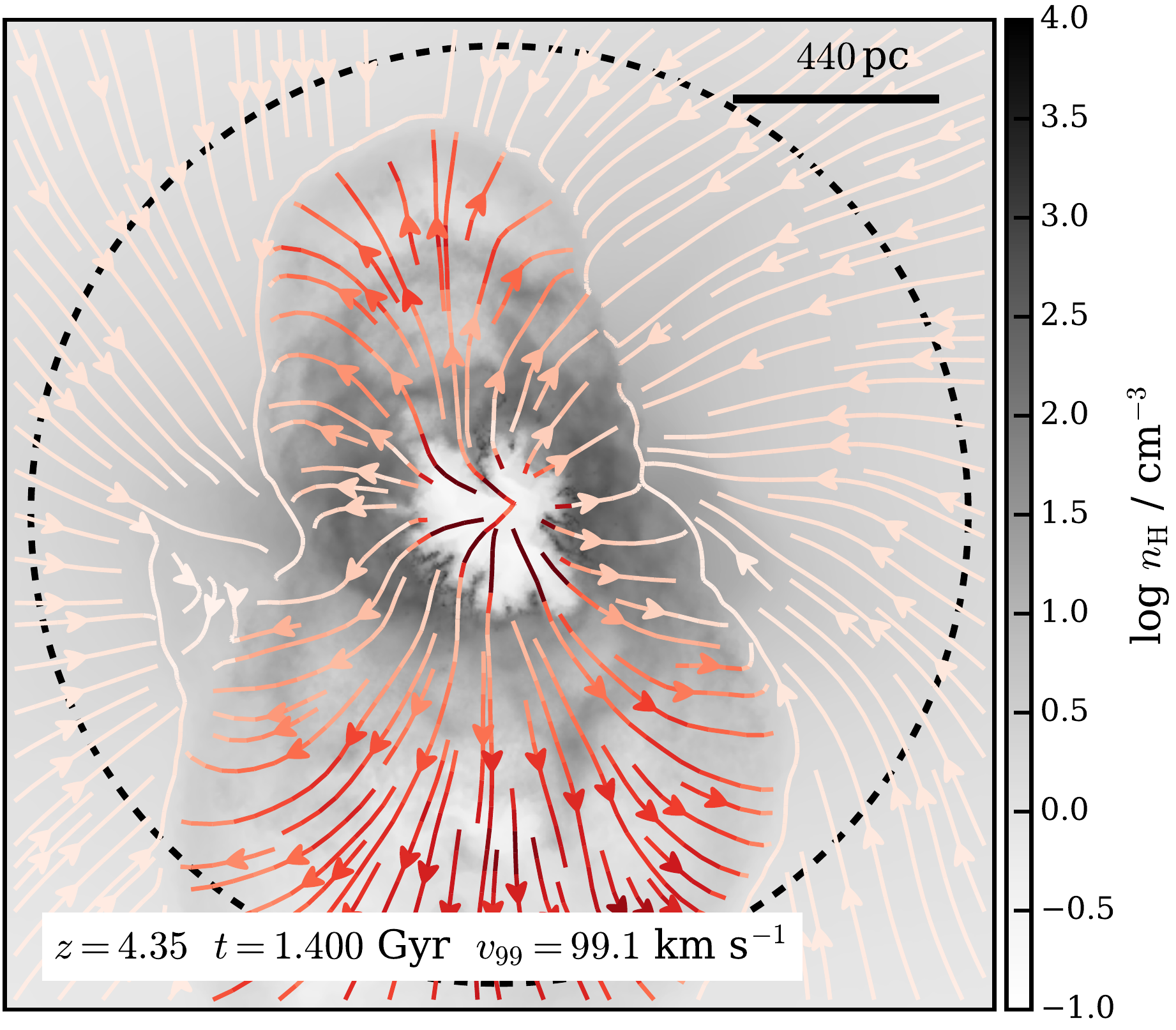}
  \includegraphics[width=0.45\textwidth]{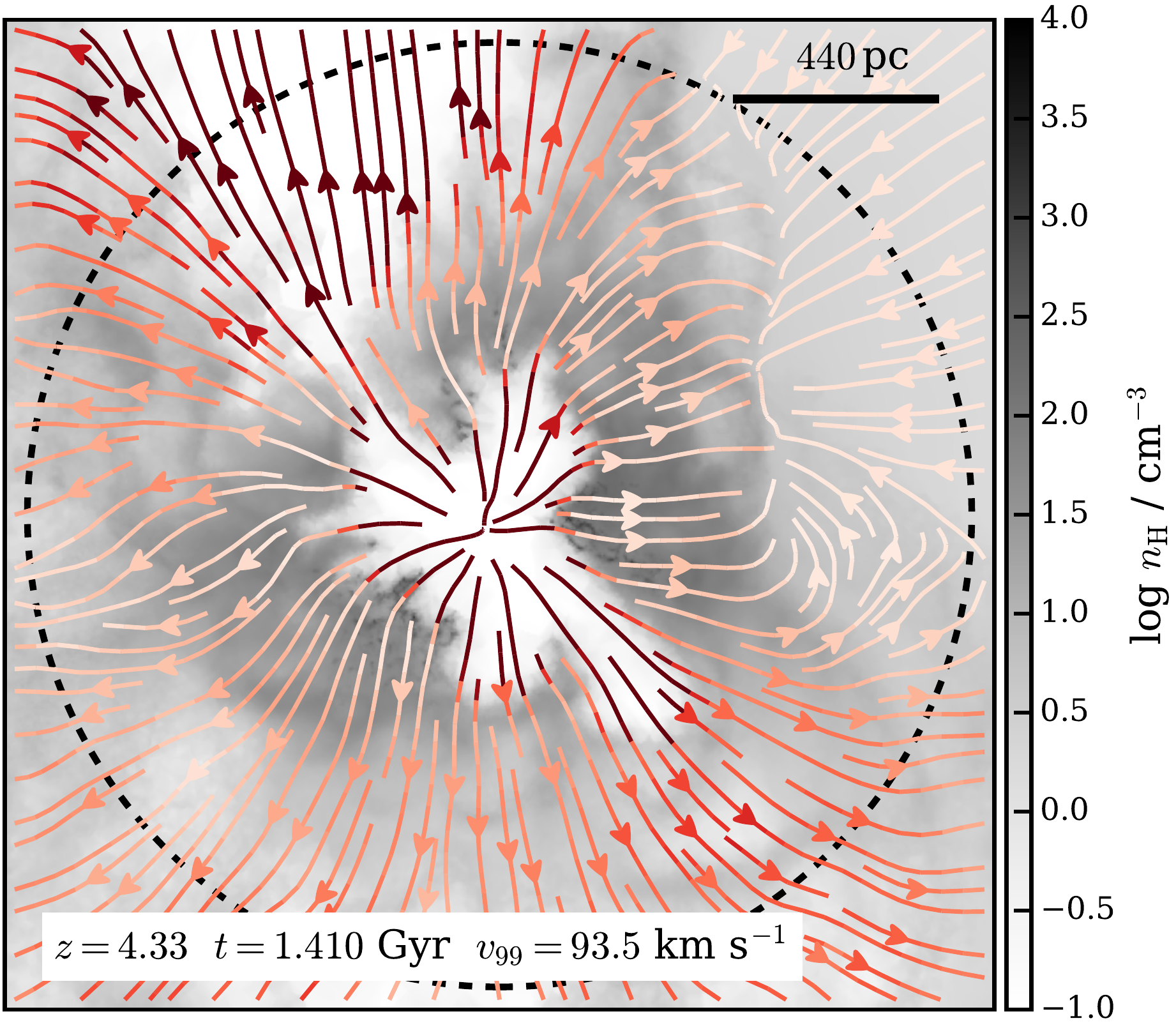}
  \includegraphics[width=0.45\textwidth]{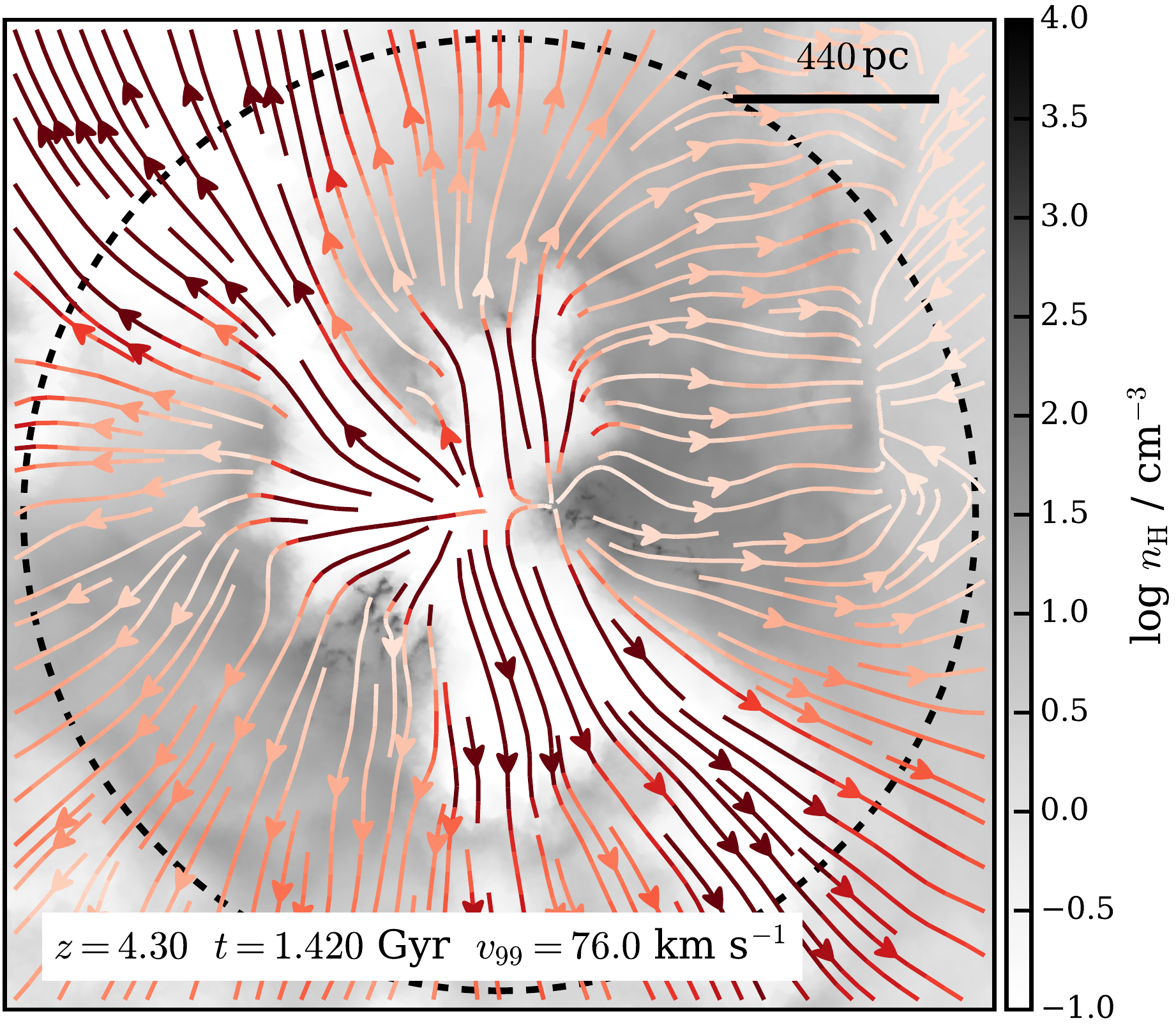}
  \caption[]{Six snapshots from the ``homogeneous'' model, each 10\,Myr apart, starting at $t=1.37$\,Gyr (as read from left to right). The grey scale map encodes the projected density. Colored stream lines show the projected $xy$-velocity field. The darker the stream lines, the higher the velocity. Each panel re-normalizes the velocity coloring such that the darkest red is $v_{99}$, the 99th percentile of velocities. This value is shown at the bottom of each panel. {{$0.25\,\Rvir$ is indicated as a dashed circle.}} Initially, in the first panel, there is inflowing gas accreting from multiple directions. Then, in the fourth panel, a super bubble begins expanding, further strengthened by additional supernova explosions in the third panel. Within 30-40 Myr these flows reach the virial radius.}
  \label{fig:stream}
\end{figure*}
%%%%%%%%%%%%%%%%%%%%%%%%%%%%%%%%%%%%%%%%%%%%%%%%%%%%%%%%%%%%%%%%%%%%%%%%%%%
%%%%%%%%%%%%%%%%%%%%%%%%%%%%%%%%%%%%%%%%%%%%%%%%%%%%%%%%%%%%%%%%%%%%%%%%%%%
\begin{figure}
  \includegraphics[width=\linewidth]{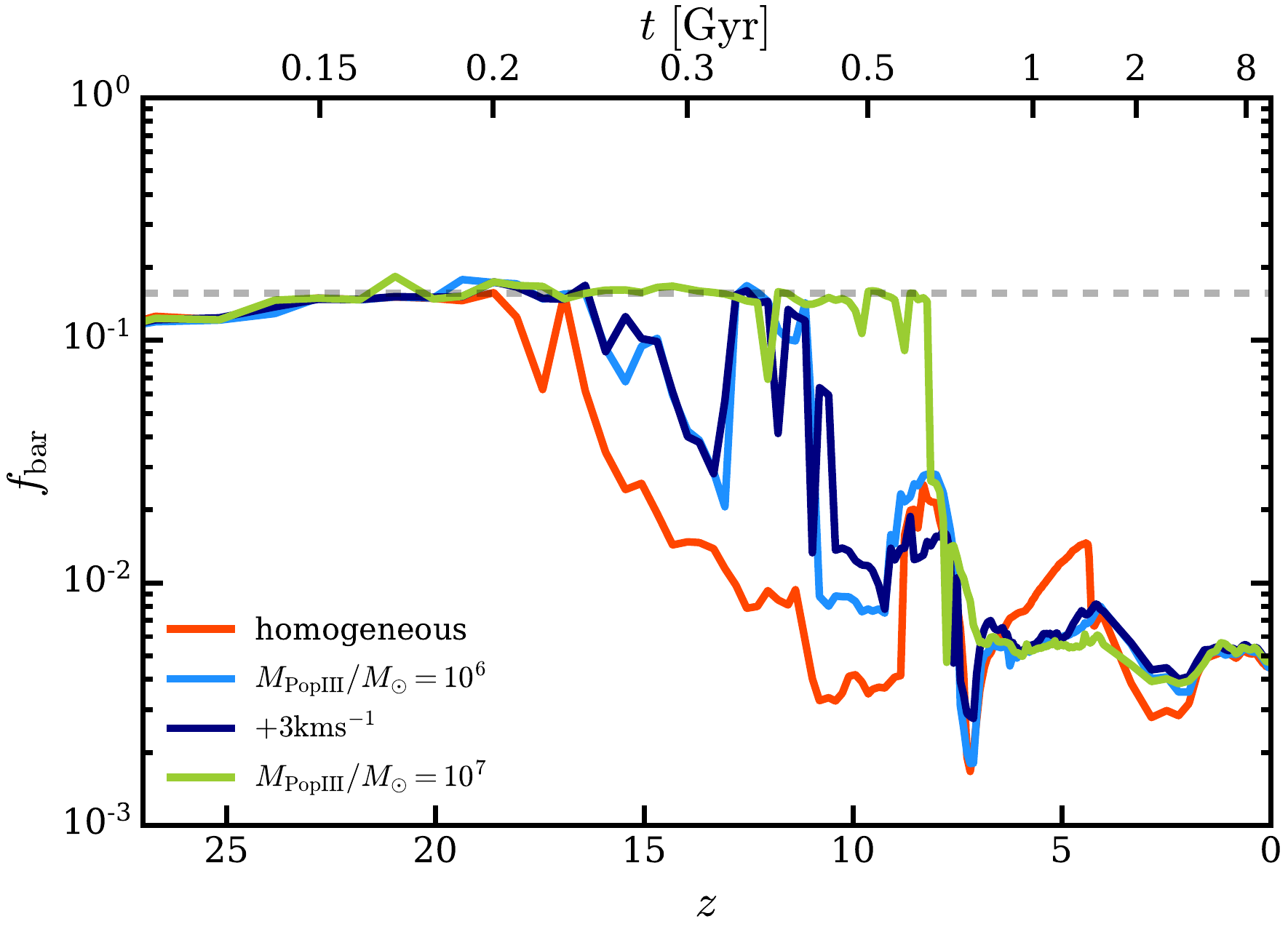}
  \caption[]{Fraction of baryons within the virial radius as a function of redshift. The grey dashed line shows the cosmic baryon fraction of 0.156. The low conversion efficiency of dwarf galaxies is already determined around $z\approx8$ as they have expelled most of their gas. The precise timing depends on the onset of star formation, which varies between the three PopIII models.}
  \label{fig:fbar}
\end{figure}
%%%%%%%%%%%%%%%%%%%%%%%%%%%%%%%%%%%%%%%%%%%%%%%%%%%%%%%%%%%%%%%%%%%%%%%%%%%

\subsection{Halo properties}
\label{sec:haloprops}
We present a cosmological zoom simulation of a $2\times10^9\Msun$ halo mass dwarf galaxy. The initial conditions were created by \cite{Jenkins2013} and are in a sub-box derived from the EAGLE \citep{Schaye2015} simulation volume, chosen to feature an isolated system. The full box considered here has a side length of  $L=12.5\,h^{-1}\mathrm{Mpc}$ and includes three zoom levels. {{The highest resolution level contains the entire Lagrangian region of the $z=0$ system. This means that all the matter that collapses to form the final galaxy is resolved at the highest level at all times. In this region,}} the dark matter mass is $m_\mathrm{DM}\approx 80\Msun$ with $\varepsilon_\mathrm{DM}=10\,$pc gravitational softening length that is held fixed at its physical value from $z=10$ to $z=0$. We set the target gas mass resolution to $m_\mathrm{gas}=4\Msun$ (which is kept fixed by appropriate local refinement and derefinement operations) with a gravitational softening of $\varepsilon_{\rm gas}=0.5\,$pc. The stellar softening is set to $\varepsilon_\star=4\,$pc. {{Beyond the Lagriangian region, the density and, thus the long-range gravitational force, is sampled only by dark matter. Each zoom level outwards increases the dark matter particle mass and its softening by a factor of 8.}} We will compare to observations where appropriate, and attempt to determine how similar our simulated dwarf galaxies are to Local Group dwarf galaxies despite the fact that the simulation is chosen to be an isolated system.

This same initial condition is run three times, varying the setting for PopIII enrichment. The first ``homogeneous'' calculation begins with all gas at a metallicity of [Fe/H]$ = -4$, whereas ``$10^6$'' and ``$10^7$'' begin with metal-free gas. Then, with the aid of the on-the-fly halo finder ({\small Subfind}, \citealt{Springel2001}) the gas in each halo that crosses the mass of $M_{\rm PopIII}$ is enriched to [Fe/H]~$ = -4$ \citep{Bromm2001}. Gas cooling is disabled below [Fe/H]~$ = -4$. Lastly, we run one additional simulation ``\tkms'' that is identical to ``$10^6$'' except that each star is given an additional small velocity kick at birth. The direction of the kick vector is random, and the magnitude is fixed to $3\,\mathrm{km\,s}^{-1}$. This run aims to compensate for unresolved stellar motion below the softening scale. This motion may originate from turbulence in the star-forming cloud cores or from two-body interactions between stars in a dense cluster. We remain agnostic about the process and instead set the value to match the measured velocity dispersion in star clusters \citep[e.g.][]{Gieles2010, Kuhn2019, Theissen2021}.

We first characterize the general properties of the halo. In the upper panel of Fig.~\ref{fig:SMHM} we show the the total stellar mass in \Rvir as a function of total mass in \Rvir at $z=0$. We include the various extrapolated results from abundance matching by \cite{Behroozi2013}, \cite{Moster2013} and \cite{Guo2011} as solid, dot-dashed and dotted lines, respectively. Our $z=0$ final stellar masses are between $7\times10^5 - 2\times10^6$\,\Msun~in the four models. This range of values is well within expectations for the final halo mass of $2\times10^9$\,\Msun. We additionally show the five EDGE halos presented in \cite{Orkney2021} as blue circles, and the FIRE-2 dwarfs presented in \cite{Wheeler2017}. It is interesting to note that our simulations lie in the same region as the EDGE simulations, whereas the FIRE-2 models exhibit significantly lower stellar masses. Both EDGE and \lyra have similar implementations of SN feedback, resolving the Sedov radii of individual bursts with a pure thermal injection. FIRE-2 on the other hand has an explicit momentum feedback that seems to result in a much stronger reduction of SF.
%While being within constraints, our stellar masses are on the high side. 
We note that our feedback prescription does not yet include other energetic sources besides supernovae. Adding other processes known be to relevant, such as stellar winds, radiation from young stars, photo-electric heating, or magnetic fields and cosmic rays, may decrease the final stellar mass. Note that the differences in our PopIII parametrization do not affect the final halo mass but do have a small effect on the final stellar mass. 

The lower panel of Fig.~\ref{fig:SMHM} shows the velocity dispersion of the halo as a function of redshift for each model. While there is some variation at early times ($15>z>8$), during the star forming phase, all four models end up following the same evolution and exhibit the same $z=0$ value of $\sigma=12.3\,\mathrm{km\, s}^{-1}$. However, in Fig.~\ref{fig:rotcurve} we see that the different models have a strong effect on the central region, as evidenced by the rotation curves split into the DM, stellar and gaseous components. The largest difference arises from the varying central concentration of the stars. ``$10^7$'' forms a large stellar core that dominates the central potential out to $\gtrsim 50\,$pc. The stellar component of this simulation was formed hierarchically from larger halos. Each of these progenitors formed its stars in a large monolithic collapse after a PopIII enrichment event occurred. The ``$10^6$'' model, on the other hand, formed from a larger number of smaller halos, each of which had a lower central stellar density. Thus, the region dominated gravitationally by the stars only extends to $\sim 20\,$pc. As we will see in the following sections, this difference has a number of important consequences.

%%%%%%%%%%%%%%%%%%%%%%%%%%%%%%%%%%%%%%%%%%%%%%%%%%%%%%%%%%%%%%%%%%%%%%%%%%%
\subsection{Star formation and quenching}
\label{sec:sf}
%%%%%%%%%%%%%%%%%%%%%%%%%%%%%%%%%%%%%%%%%%%%%%%%%%%%%%%%%%%%%%%%%%%%%%%%%%%

To obtain a better understanding of the evolution of this system, we will now describe the star formation and subsequent quenching of our galaxy. Fig.~\ref{fig:sfr} displays the star formation history (SFH) for our three dwarfs. We do not show the SFH of ``\tkms'' since it is identical to the ``$10^6$'' model. The measured SFH is based on all the stars that are within the virial radius at $z=0$,  so while many stars form ex-situ, they are included in the final SFH of the whole halo. The SFH is characterized by one main SF phase between $z\approx20-8$. Peak rates of $1-5\times10^{-2}\,\mperyr$ are reached for short periods of time and then rapidly decline. The SF phase comes to an end in all four simulations at the onset of reionization around $z\approx7.5$. Then at around $z\approx4.5$, the galaxy experiences a merger with a mass ratio of approximately 1:10. Two of the simulations exhibit a second burst of SF in response to the merger, while ``$10^6$'' and ``\tkms'' do not. Even between the two models that do restart forming stars, the SFRs during this burst differ by more than two orders of magnitude. This is a further manifestation of the effects of the different PopIII enrichment modes. 

The onset of SF is delayed by a few hundred megayears in the two inhomogeneous models. This is due to their later enrichment. In the ``homogeneous'' model, the first star forms at $z=19.5$, while in ``$10^6$'', the first halo is enriched at $z=17.4$ and the first star forms at $z=16.9$. Finally, in ``$10^7$'', the first halo is enriched at $z=12.8$ and the first star forms at $z=12.6$. These delays allow the halos to grow and accumulate more gas. When the gas is finally enriched, the ``$10^7$'' model produces a large peak in SF as gas suddenly collapses monolithically and forms many stars simultaneously. This in turn causes a SF valley directly afterwards before self-regulation sets in. 

The ``$10^6$'' model shows a different behaviour. As there are many more $10^6$\Msun halos, and they appear continually, the SF commences more gradually and retains a fairly constant rate up until the galaxy is quenched. This model forms more stars in a shorter time than the ``homogeneous'' model.
% Central density doesn't reach same value in the different simulations.
 We note that the stellar mass formed after $z=4$ contributes $<1\%$ to the total stellar mass in all four models. {{This implies that the cumulative SFH of our model galaxy is consistent with the ``Oldest'' model considered in Fig.~9 of \cite{Weisz2014}. Among the observations of LG dwarfs, certain dwarf spheroidals such as Draco and Sculptor show evidence of such ancient stellar populations. So, while it is rare to find little to no ongoing SF in the LG sample, such a SFH is by no means ruled out.}}
 
%  Dwarf galaxies in the mass regime we at looking at are often quenched at high redshift. The quenching occurs through a combination of heating from the UV background and violent feedback that heats and ejects large fractions of gas from the centers. Both these processes are able to quench dwarf galaxies due to the low mass and low virial temperature.
 
% The SFH is characterized by extreme burstiness.  Peak periods are repeatedly interspersed with periods of no star formation. Each peak period forms enough stars that multiple SNe explode within a short period of time, thus creating a super-bubble that drives hot outflows and reduces the SF for some hundreds of megayears. If sufficient gas remains, it will re-collapse and produce a subsequent peak.

Many authors have investigated the high redshift quenching of dwarf galaxies due to the onset of reionization {{using simulations}} \citep[e.g.][]{Onorbe2015,Munshi2013, Wheeler2019, Agertz2020}.  During reionization, the ultra-violet (UV) background increases and reaches its peak values around $z=8-7$. In small halos {{($\Mhalo\lesssim5\times10^9\Msun$)}} the central gas densities are too low for the cold gas to become self-shielded. Instead, the UV radiation sweeps through and heats the gas above the virial temperatures of these halos, such that they become quenched. This picture is corroborated by observations of LG dwarf galaxies. \cite{Weisz2014} model the SFH of various dwarfs and show that the majority of their stars are ancient. Nevertheless, certain objects have small populations of younger stars or even ongoing star formation at low rates. 

In slightly more massive halos, where the density is sufficiently large, self-shielding can protect the cold gas from this radiation. Star formation can thus proceed beyond the time of reionization. However, if the mass of the self-shielded gas is small enough, a single SN can be sufficient to blow apart the self-shielded clump and allow the radiation to enter. Then, even these galaxies become quenched. This is the case with our simulated galaxy, although it requires multiple SNe to shut down SF. 

To illustrate this, we show the outflow (in the ``homogeneous'' model) caused by the SF episode during the merger at $z\approx4.5$ in Fig.~\ref{fig:stream}. Each panel represents a snapshot spaced $\sim10$\,Myr apart. To make the outflowing behaviour clear, we overlay the gas density (gray colormap) with the velocity field (arrows colored by their velocity, redder is faster). As the outflows increase their velocity, the 99-th percentile velocity, $v_{99}$, grows. $v_{99}$ is given at the bottom of each panel and sets the darkest color of the arrows. The virial velocity of the galaxy at this point is $\sim13.5\,{\rm km\, s}^{-1}$. 

As we can see in the top two panels, there is a steady inflow at first as the 1:10 merger enters the halo. The inflow is centered on the small incoming galaxy rather than on the central halo, since the central galaxy is currently quenched. In the third panel, the signature of a SN shell is visible in the central gas density. The merger has occurred and SF has commenced. While the velocity field is here still dominated by inflows, the center is slightly perturbed. $10\,$Myr later the SN bubble has increased rapidly in size and has already reached the virial radius (dashed line) in the south. The peak velocities occur in the outflows, and the inflows are halted towards the center as they meet the outflows. The final two panels see an increase in the volume coverage of the outflows. Almost all inflow within $R_{200}$ has ceased, and the central gas density of the halo is $10^{-1}\mathrm{cm}^{-3}$ and lower. SF remains arrested after this episode of outflowing gas.

\subsubsection{Outflows and mass loading}
To gain a quantitative understanding of the differences in the outflows between the four models, Tab.~\ref{tab:loading} shows the median mass loading ($\eta_\mathrm{M}$), metal mass loading ($\eta_\mathrm{Z}$) and energy loading ($\eta_\mathrm{E}$) factors. The detailed description of how these values are calculated is given in \cite{Gutcke2021}. These values depend on $\dot{M}_\star$, so they are only non-zero during the SF phase of each galaxy.  They are intended as a fractional measure of how much mass, metal mass and energy are transported out of the galaxy relative to what is injected by the SNe. Interestingly, the \tkms~ model displays mass and metal mass loading factors increased by around 0.5 dex (and 0.25 dex for the energy loading) compared to ``$10^6$''. The additional velocity kicks produce ``run away'' stars \citep[i.e.][]{Andersson2021} that move out of their natal clouds to explode as SNe in lower density material. SNe in low density environments are more efficient at entraining gas and driving it out of the disk \citep[i.e.][]{Gatto2015}.

It is also apparent that the ``$10^7$'' model is an outlier in all of its loading factor values. The delayed onset of SF and subsequent monolithic collapse of the halo gas in this model cause a higher central stellar density (cf. Fig~\ref{fig:rotcurve}). This in turn generates more centrally-concentrated (and therefore more clustered) SNe that drive two orders of magnitude more mass and metals in the outflows. 

This is substantiated in Fig.~\ref{fig:fbar}, where we show the baryon fraction within $R_{200}$ as a function of redshift. The ``homogeneous'' model shows the slowest decline in baryons (the mass loading factor is also smallest in this model). ``$10^7$'' keeps the cosmic baryon fraction almost until reionization and then the value plummets to meet the other models. However, these differences are minor relative to cosmic time. All four models have their low fractions set before $z=6$ which is in the first gigayear. The values remain low for the next $13$~Gyr until $z=0$. This differs from the predictions made by \cite{Onorbe2015} who found that the fraction decreases steadily and continually across time. As SF is also shut down by reionization in their model, this distinction is likely caused by the different implementation of feedback and the ensuing outflows.

We have now characterized the high redshift behaviour of our galaxies. However, the only observations of dwarfs in this mass range are in the vicinity of the Milky Way. Thus, we will now turn to a $z=0$ comparison of the properties of our galaxies and observations of Local Group dwarfs.

%%%%%%%%%%%%%%%%%%%%%%%%%%%%%%%%%%%%%%%%%%%%%%%%%%%%%%%%%%%%%%%%%%%%%%%%%%%
\begin{table}
	\begin{center}
		\begin{tabular}{|l c c c|}
			\hline\hline
			Name & $\eta_\mathrm{M}$ & $\eta_\mathrm{Z}$ & $\eta_\mathrm{E}$ \\
			\hline
			homogeneous & $0.74$ & $-1.68$ & $-1.58$ \\
			$M_{\mathrm{PopIII}}/M_{\odot}=10^6$ & $0.65$ & $-1.88$ & $-1.27$ \\
			$+3~\mathrm{km~s}^{-1}$ & $1.18$ & $-1.35$ & $-1.01$ \\
			$M_{\mathrm{PopIII}}/M_{\odot}=10^7$ & $2.44$ & $0.84$ & $0.24$ \\
			\hline\hline
		\end{tabular}
	\end{center}
	\caption{Median values for the mass, metal mass and energy loading factors for each simulation. ``$10^7$" has 1-2 orders of magnitude stronger outflows than the other three simulations. }
	\label{tab:loading}
\end{table}
%%%%%%%%%%%%%%%%%%%%%%%%%%%%%%%%%%%%%%%%%%%%%%%%%%%%%%%%%%%%%%%%%%%%%%%%%%%

%%%%%%%%%%%%%%%%%%%%%%%%%%%%%%%%%%%%%%%%%%%%%%%%%%%%%%%%%%%%%%%%%%%%%%%%%%%
\begin{figure}
  \includegraphics[width=\linewidth]{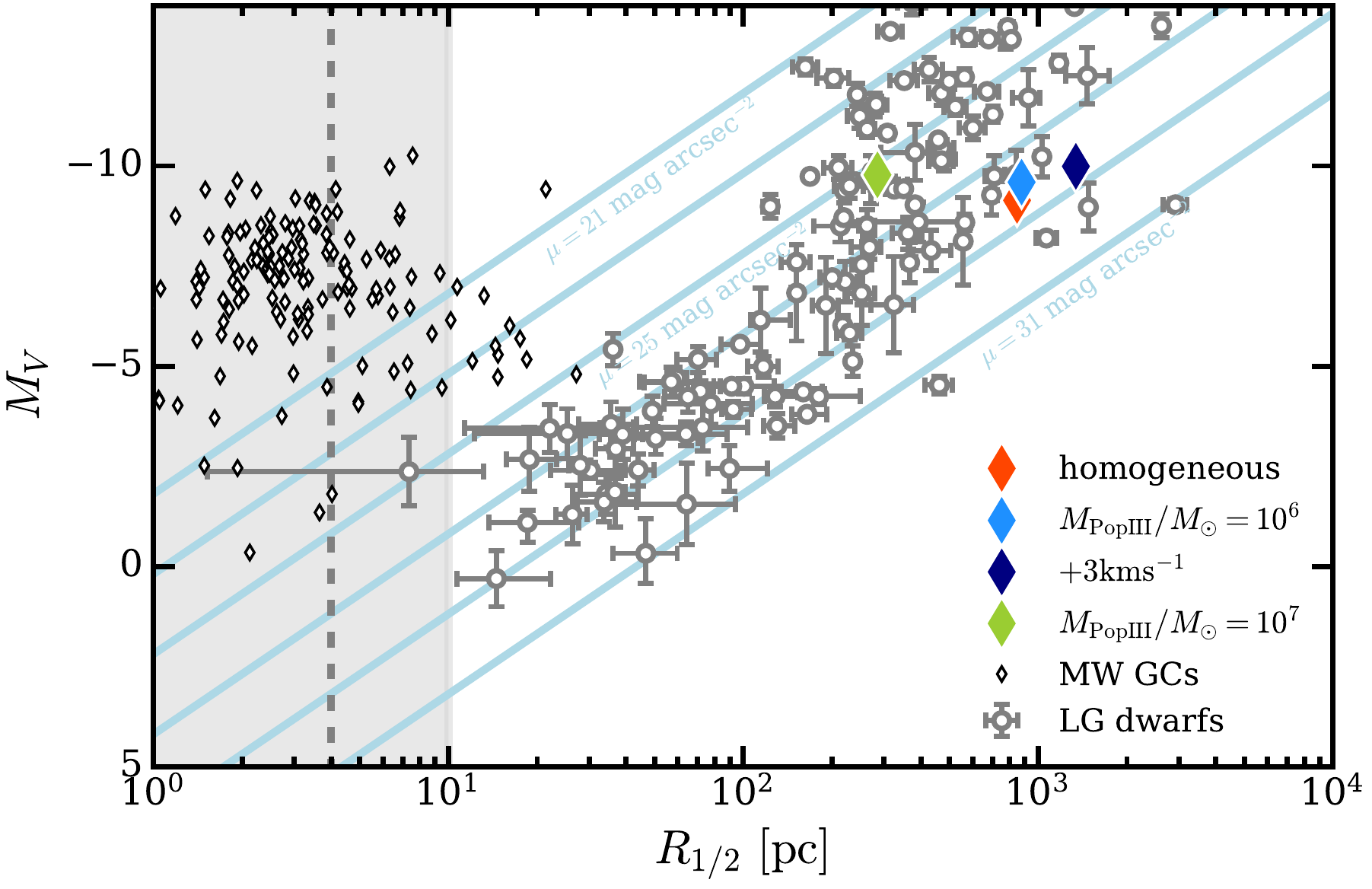}
  \caption[]{Size - magnitude relation showing observational data from the compilation of \cite{McConnachie2012}. Globular cluster data is taken from the updated Milky Way cluster catalogue from \cite{Harris1996} (2010 edition). The grey band indicates the area below our DM softening length. ``$10^7$'' is again decidedly different from the other three simulations. Its smaller size at approximately the same magnitude is caused by the larger central stellar bulge, or ``nuclear star cluster''.}
  \label{fig:size-mass}
\end{figure}

%%%%%%%%%%%%%%%%%%%%%%%%%%%%%%%%%%%%%%%%%%%%%%%%%%%%%%%%%%%%%%%%%%%%%%%%%%%
\subsection{Validation of $z=0$ properties}
\label{sec:z0}
%%%%%%%%%%%%%%%%%%%%%%%%%%%%%%%%%%%%%%%%%%%%%%%%%%%%%%%%%%%%%%%%%%%%%%%%%%%
\subsubsection{Size--magnitude relation}
%%%%%%%%%%%%%%%%%%%%%%%%%%%%%%%%%%%%%%%%%%%%%%%%%%%%%%%%%%%%%%%%%%%%%%%%%%%
To validate our simulation against LG observations, we first turn to the size-magnitude relation. Fig.~\ref{fig:size-mass} shows the stellar half light radius versus the absolute V-band magnitude at $z=0$. Local Group dwarfs \citep[compiled by][]{McConnachie2012} are shown as gray circles with error bars, Milky Way globular clusters (\citealt{Harris1996}, 2010 edition) as black diamonds. The $z=0$ stellar half light sizes are around $R^\star_{1/2}\approx 850\,$pc, except for ``$10^7$'' which has a radius of $R^\star_{1/2}= 285\,$pc. Since the majority of stars formed in the first $2$\,Gyr of cosmic time ($z\sim10-4$), the stellar population is very old and faint with an absolute V-band magnitude of $M_V\approx-9$.

The location of our simulations on the size-magnitude relation puts our dwarfs into the classical dwarf regime rather than into the ultra-faint regime. This is interesting, since previous simulation work on dwarf galaxies generally predict the dark matter halos of ultra-faint dwarfs to be around $10^9\Msun$ (e.g. \citealt{Wheeler2019, Onorbe2015}), which is only slightly lower than our halo mass. This difference is also apparent in the location on the \Mstar-\Mhalo plane (Fig.~\ref{fig:SMHM}), where our simulations host significantly more stellar mass per dark matter mass. However, since the abundance matching relation is not well constrained in this low-mass regime, neither is ruled out by it. Indeed, it is likely that there is a large amount of scatter in this regime, and that the same halo mass may support orders of magnitude differences in stellar mass depending on the environment and specific history of each dwarf. 

But we may summarize here that our four models all match the observations well. The reason that the ``$10^7$'' model is different from the other three is that its half light radius is significantly smaller. This in turn is due to a more pronounced central stellar bulge, which we tentatively identify as a nuclear star cluster. The detailed nature of the stellar morphology will be the subject of a future investigation.

%%%%%%%%%%%%%%%%%%%%%%%%%%%%%%%%%%%%%%%%%%%%%%%%%%%%%%%%%%%%%%%%%%%%%%%%%%%
\begin{figure}
  \includegraphics[width=\linewidth]{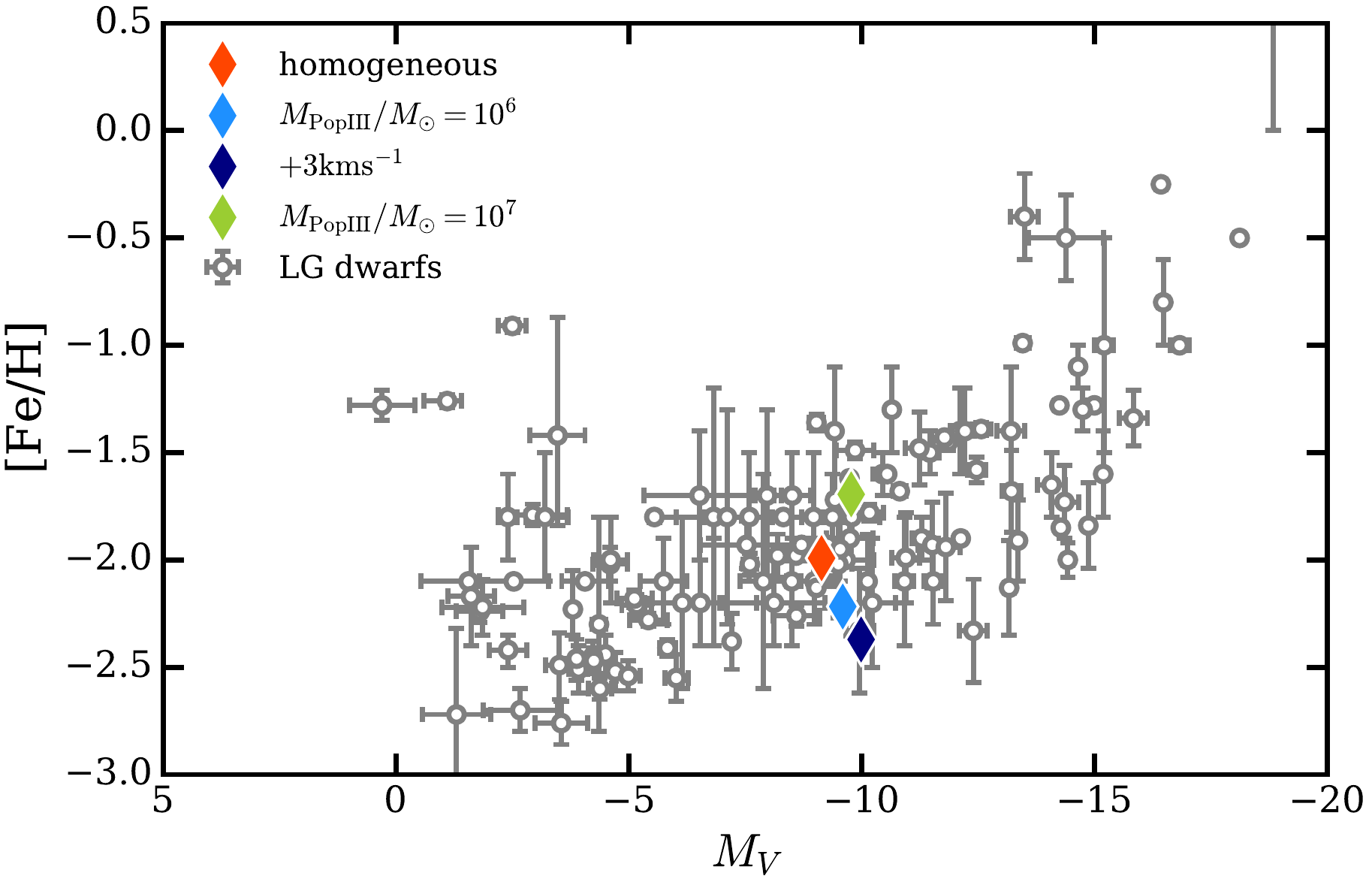}
  \caption[]{Absolute V-band magnitude versus metallicity with the $z=0$ values for our cosmological simulations. Observational data are from the compilation of nearby dwarf galaxies in \cite{McConnachie2012}. Luminous galaxies show a positive relation with metallicity. However, the trend flattens towards galaxies fainter than $M_V\approx-12$ around a metallicity of [Fe/H]~$\lesssim-2$. Our simulations sit well within the scatter of this flatter tail.}
  \label{fig:mv-metallicty}
\end{figure}
%%%%%%%%%%%%%%%%%%%%%%%%%%%%%%%%%%%%%%%%%%%%%%%%%%%%%%%%%%%%%%%%%%%%%%%%%%%
%%%%%%%%%%%%%%%%%%%%%%%%%%%%%%%%%%%%%%%%%%%%%%%%%%%%%%%%%%%%%%%%%%%%%%%%%%%
\begin{figure*}
  \includegraphics[width=\linewidth]{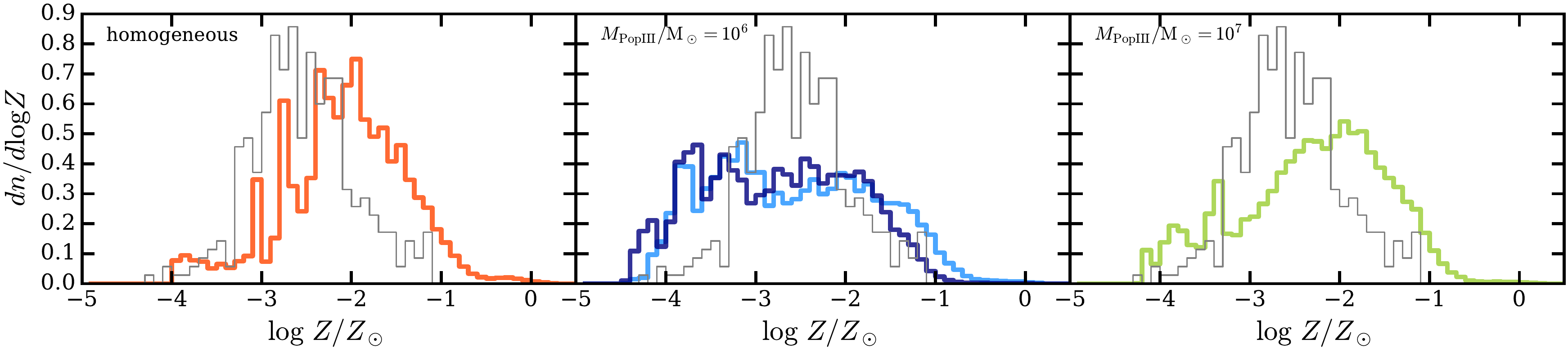}
  \caption[]{Metallicity distribution function (MDF) of stars within the half-mass radius (normalized per bin in metallicity) for each simulation. From left to right: the ``homogeneous'', ``$10^6$'' and ``$10^7$'' models. The thin solid black line in each panel shows the compiled MDF of individual stars in LG dwarf galaxies taken from \cite{Simon2019}. }
  \label{fig:mdf}
\end{figure*}
%%%%%%%%%%%%%%%%%%%%%%%%%%%%%%%%%%%%%%%%%%%%%%%%%%%%%%%%%%%%%%%%%%%%%%%%%%%
%%%%%%%%%%%%%%%%%%%%%%%%%%%%%%%%%%%%%%%%%%%%%%%%%%%%%%%%%%%%%%%%%%%%%%%%%%%
\begin{figure*}
  \includegraphics[width=\linewidth]{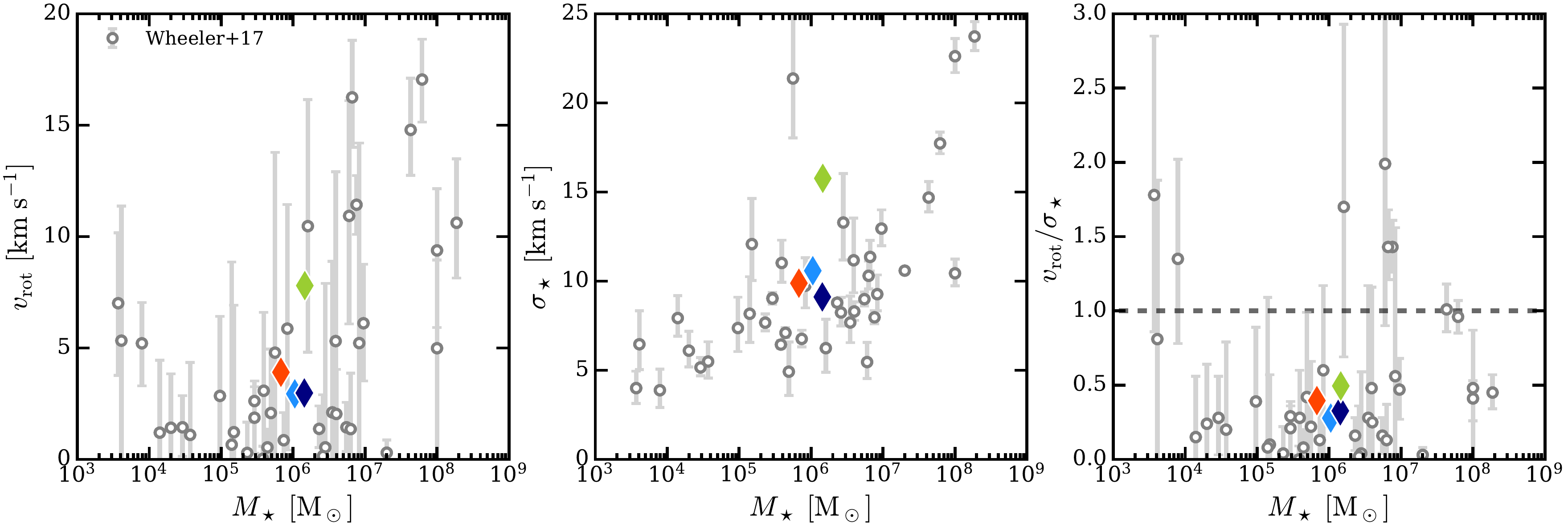}
  \caption[]{$v_\mathrm{rot}$, $\sigma_\star$ and their ratio as a function of stellar mass. Colored diamonds show the four simulations at $z=0$. Grey circles with errors are the observations from \cite{Wheeler2017}. As can be seen in the right-most panel, our simulations are dispersion dominated, in line with expectations from dwarfs of similar mass.}
  \label{fig:kinematics}
\end{figure*}
%%%%%%%%%%%%%%%%%%%%%%%%%%%%%%%%%%%%%%%%%%%%%%%%%%%%%%%%%%%%%%%%%%%%%%%%%%%
%%%%%%%%%%%%%%%%%%%%%%%%%%%%%%%%%%%%%%%%%%%%%%%%%%%%%%%%%%%%%%%%%%%%%%%%%%%
\subsubsection{Magnitude--metallicity relation}
%%%%%%%%%%%%%%%%%%%%%%%%%%%%%%%%%%%%%%%%%%%%%%%%%%%%%%%%%%%%%%%%%%%%%%%%%%%
A further strong constraint on dwarf galaxy evolution is given by the stellar mass-metallicity relation. Indeed, \cite{Agertz2020} and \cite{Prgomet2021} indicate that this relation can inform us strongly about the physics relevant in dwarf galaxies, as indicated also by the challenges it has posed for various simulation projects \citep[e.g.][]{Wheeler2019}. We show the observations from the MW and M31 satellites in Fig.~\ref{fig:mv-metallicty} in comparison to our simulations. At the magnitude of our galaxy, we obtain a stellar metallicity that is well within expectations. The metallicity value is here computed following the common practice of calculating the mean of the logarithmic values according to
\begin{equation}
  % {\rm [Fe/H]} = \frac{1}{M_{\star, \mathrm{tot}}}\sum^N_{i=0} M_i \log \Bigg(\frac{Z_{i}}{Z_{\odot}}\Bigg).
   {\rm [Fe/H]} = \frac{1}{N}\sum^N_{i=1} \log \Bigg(\frac{Z_{i}}{Z_{\odot}}\Bigg).
\end{equation}

The differences in mean metallicity between the models can be explained with a look at the metallicity distribution functions (MDFs). In Fig.~\ref{fig:mdf} we show the MDFs for each simulation in comparison to the compiled distribution of LG dwarf galaxies \citep{Simon2019}. We exclude stars $i$ whose metallicity is exactly {{$\log ({Z_{i}}/{Z_{\odot}})=-4$}} since they are at our threshold. The fraction of these stars is 33.8\%, 18.7\%, 20.0\% and 9.4\% for ``homogeneous'', ``$10^6$'', ``\tkms'' and ``$10^7$'', respectively. As we can see, even though the average metallicity of the dwarf galaxies is well reproduced, the distribution varies between the four models. The uni-modal versus bi-modal distinction can be explained well with a comparison to Fig.~\ref{fig:firstsubs}. The two models that have a clear bi-modality in their MDF have a stronger bi-modality in their distribution of first-star halo masses. Small halos forming stars will inevitably form less stars (due to gas availability) for a shorter period of time. Because SF is not sustained for long enough in these small halos, they do not provide enrichment for multiple generations of ever more enriched stars. Thus, the metallicity of each star is lower, giving rise to a low-Z peak ($\log ({Z_{i}}/{Z_{\odot}})\approx -4$) in the MDF. Larger halos will form more stars for a longer period of time, allowing new stars to emerge from recently enriched gas, thus giving rise to the high-Z peak ($\log ({Z_{i}}/{Z_{\odot}})\approx -2$) in the MDF.
{{We note that the differences seen between Fig.~\ref{fig:mv-metallicty} and Fig.~\ref{fig:mdf} are due to the $\log ({Z_{i}}/{Z_{\odot}})=-4$ stars.}}
%While ``$10^6$'' and ``\tkms'' also underwent the merger at $z\approx4.5$, the density increase was not sufficient to restart star formation. The lack of late-time SF (that would produce higher metallicity stars) is what sets the overall lower metallicity in these two models. ``$10^7$'' has a higher metallicity as it formed the most stars between $z=0.2 - 0$. However, none of these differences are sufficient to favor one model over another.
We additionally note that the strong outflows allow many metals to escape. The metal escape fractions can also have a strong influence on the MDFs and the mean metallicities.

Despite the broad agreement, we note that none of the models match the observations perfectly, being generally wider or bi-modal rather than uni-modal around a value of [Fe/H]$\approx-2.8$. This may in part be due to the fact that the observational distribution shows individual stars compiled from various dwarf galaxies that have different mean metallicities {{and a distribution of stellar masses}}. This may artificially create a more centrally-concentrated distribution. Furthermore, we note that we do not yet model various physical processes such as radiation and SNIa. These processes are known to enhance the relative amount of iron, to contribute to driving turbulence, and to enhance metal mixing. Including these processes (planned for future work) will likely affect the stellar metallicity distribution by making it more unimodal.
%%%%%%%%%%%%%%%%%%%%%%%%%%%%%%%%%%%%%%%%%%%%%%%%%%%%%%%%%%%%%%%%%%%%%%%%%%%
% \begin{figure*}
%   \includegraphics[width=\linewidth]{images/stellar_Z_profile_comp.pdf}
%   \caption[]{Mean metallicity profile at $z=0$ for each simulation. }
%   \label{fig:Zprofile}
% \end{figure*}
%%%%%%%%%%%%%%%%%%%%%%%%%%%%%%%%%%%%%%%%%%%%%%%%%%%%%%%%%%%%%%%%%%%%%%%%%%%

%%%%%%%%%%%%%%%%%%%%%%%%%%%%%%%%%%%%%%%%%%%%%%%%%%%%%%%%%%%%%%%%%%%%%%%%%%%
\subsubsection{Stellar kinematics}
%%%%%%%%%%%%%%%%%%%%%%%%%%%%%%%%%%%%%%%%%%%%%%%%%%%%%%%%%%%%%%%%%%%%%%%%%%%

We additionally examine the kinematic signatures of the stars in our simulations. In Fig.~\ref{fig:kinematics}, we show $v_\mathrm{rot}$, $\sigma_\star$ and their ratio. We take $v_\mathrm{rot}$ to be the peak of the stellar rotation curve. $\sigma_\star$ is the stellar velocity dispersion measured within $r_{\star\mathrm{max}}$, the radius at which the stellar rotation curve is equal to $v_\mathrm{rot}$. In each panel, we also show the Local Group dwarf observations from \cite{Wheeler2017}. We are well within the scatter of the observations. The right-most panel shows the ratio between $v_\mathrm{rot}$ and $\sigma_\star$. This gives a measure of whether the stars are rotation or dispersion-supported. As our galaxies are well below $v_\mathrm{rot}/ \sigma_\star \textbf{}= 1$, the stars at $z=0$ can be assumed to be supported by dispersion. This in in line with expectations for LG dwarfs with ancient stellar populations.

Again, ``$10^7$'' is a slight outlier. Both its $v_\mathrm{rot}$ and $\sigma_\star$ values are higher than in the other three models. This is not sufficient to rule this model out, since the deviation is still well within the scatter of the observations. This difference is due to the smaller size and higher central stellar density in this simulation.

In summary, the $z=0$ properties of our simulations are a very good representation of LG dwarf galaxies. Their luminosities, sizes, metallicities and kinematics are well reproduced by our simulation model. {{Their morphology and estimated magnitudes place them in the morphological type dwarf spheriodal (dSph).}} Thus, with this reassuring finding we will now turn to examine some predictions of the model.

%%%%%%%%%%%%%%%%%%%%%%%%%%%%%%%%%%%%%%%%%%%%%%%%%%%%%%%%%%%%%%%%%%%%%%%%%%%
\begin{figure*}
  \includegraphics[width=\linewidth]{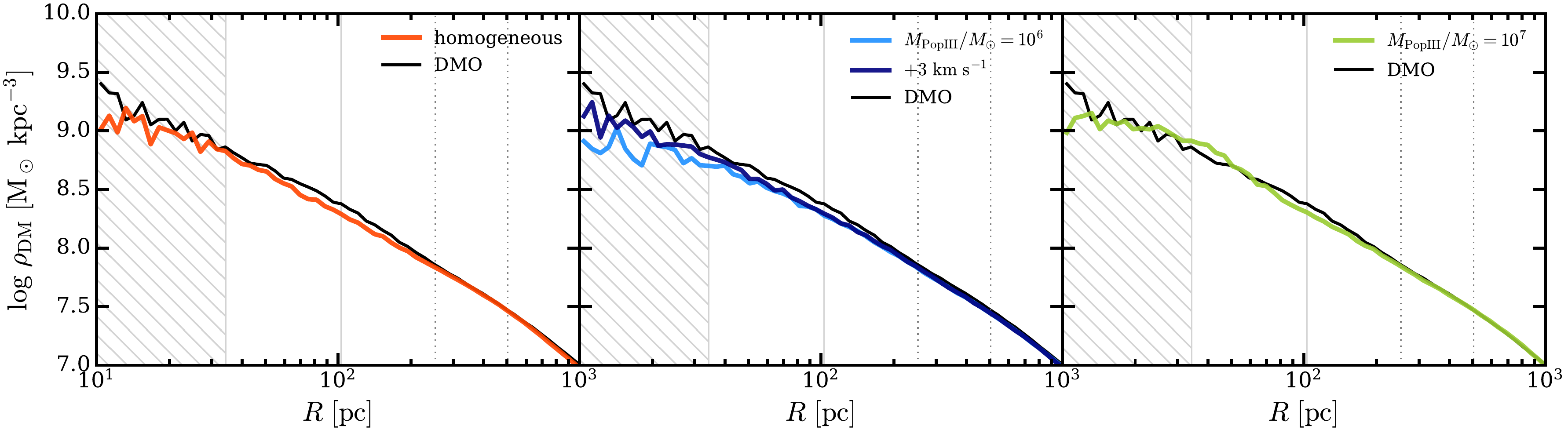}
  \caption[]{DM density profile at $z=0$ for each simulation. The black line shows the dark matter only (DMO) simulation at $z=0$. We start the $x$-axis at $\varepsilon_\mathrm{DM}=10\,$pc, the dark matter softening length. The grey shaded area extends to $r_{2000}$, the radius enclosing 2000 DM particles. The vertical solid grey line is $3\,r_{2000}$. The two vertical dotted lines indicate 1\% and 2\% of \Rvir. There is no DM core visible in the resolved part of the profile. The fits to the slope are given in Tab.~\ref{tab:dmfit}.}
  \label{fig:dmprofile}
\end{figure*}
%%%%%%%%%%%%%%%%%%%%%%%%%%%%%%%%%%%%%%%%%%%%%%%%%%%%%%%%%%%%%%%%%%%%%%%%%%%

%%%%%%%%%%%%%%%%%%%%%%%%%%%%%%%%%%%%%%%%%%%%%%%%%%%%%%%%%%%%%%%%%%%%%%%%%%%
\begin{table}
	\begin{center}
		\begin{tabular}{|l|c|c|c|}
			\hline\hline
			Name & $\alpha_{1-2\%}$ & $\alpha_{2000}$ & $\log(M_\star/M_\mathrm{halo})$ \\
			\hline
			homogen. & $-1.31  \pm  0.02$ & $-1.11  \pm  0.03$ & -3.50 \\
			$10^6$ & $-1.29  \pm  0.01$ & $-0.95  \pm  0.03$ & -3.30 \\
			$+3~\mathrm{kms}^{-1}$ & $-1.28  \pm  0.01$ & $-1.06  \pm  0.02$ & -3.17 \\
			$10^7$ & $-1.26  \pm  0.02$ & $-1.44  \pm  0.03$ & -3.16 \\
			DMO & $-1.32  \pm  0.02$ & $-1.00  \pm  0.04$ & 0 \\
			\hline\hline
		\end{tabular}
	\end{center}
	\caption{Slope of the dark matter profile measured between 1\% and 2\% \
    of the virial radius ($\alpha_{1-2\%}$). We measure it again closer to the center from $r_{2000}$, \
    the radius enclosing 2000 dark matter particles, to $3\, r_{2000}$ ($\alpha_{2000}$). We also show the stellar-to-halo mass ratio for each halo at $z=0$. The concentration of the DMO simulation is $c=15$.}
    \label{tab:dmfit}
\end{table}
%%%%%%%%%%%%%%%%%%%%%%%%%%%%%%%%%%%%%%%%%%%%%%%%%%%%%%%%%%%%%%%%%%%%%%%%%%%

%%%%%%%%%%%%%%%%%%%%%%%%%%%%%%%%%%%%%%%%%%%%%%%%%%%%%%%%%%%%%%%%%%%%%%%%%%%
\begin{figure*}
  \includegraphics[width=\linewidth]{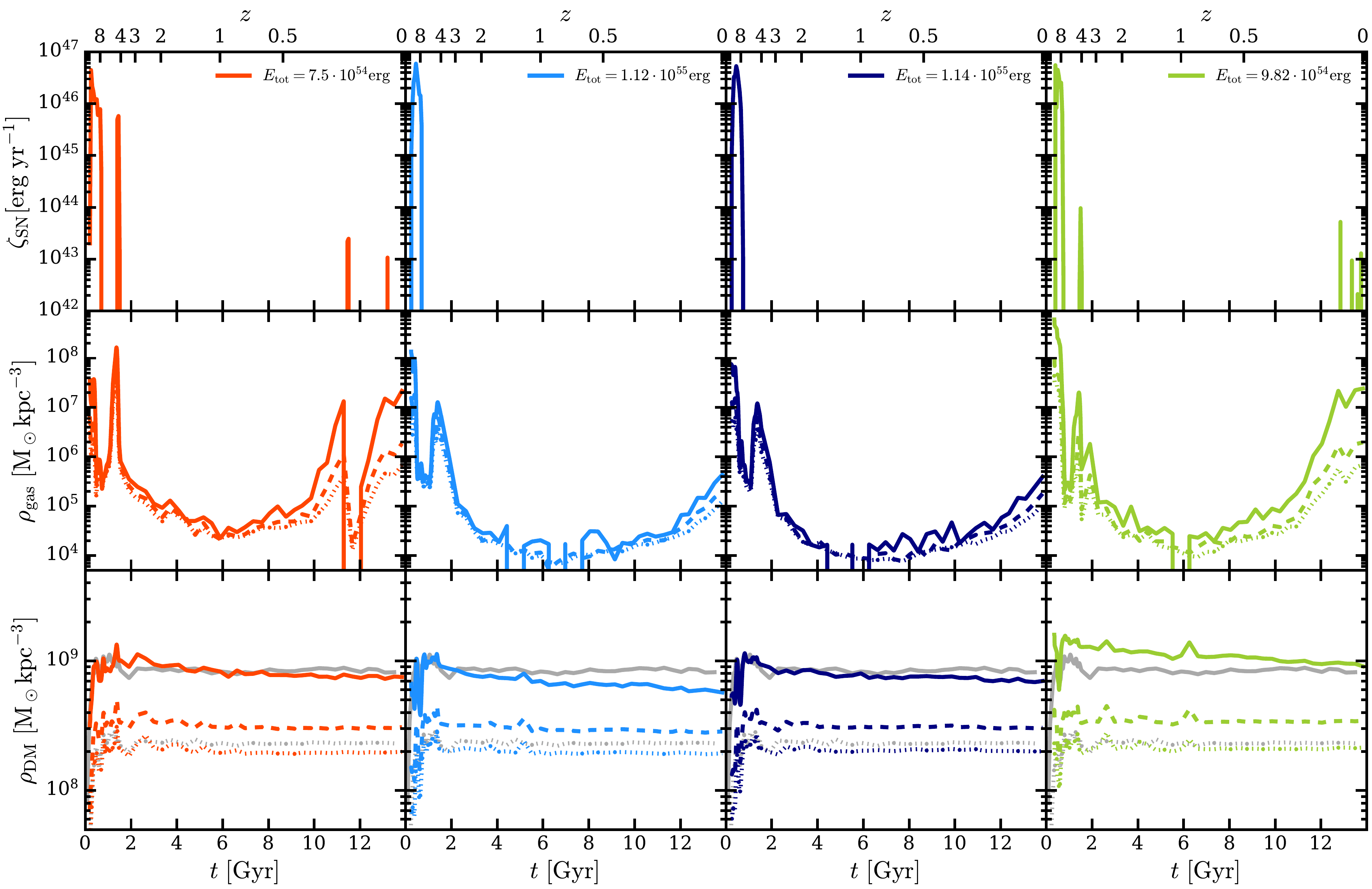}
  \caption[]{\textit{Top:} SN energy injection rate over time. \textit{Center:} Central gas density over time. \textit{Bottom:} Central dark matter density over time. In the lower two panels the solid, dashed and dotted lines are measured within $40\,$pc, $100\,$pc and $150\,$pc, respectively. The grey lines in the lower panels are the dark matter density in the DMO run, rescaled by ($1-\Omega_\mathrm{b}/\Omega_0$). Both the gas density and dark matter react to the injection of SN energy, as features match between the two. However, this is not sufficient to form a core. Slow gas accretion raises the central gas density over 6 billion years until SF is re-ignited (in two of the runs).}
  \label{fig:DM_100pc}
\end{figure*}
% \begin{figure*}
%   \includegraphics[width=\linewidth]{images/SNrate_DMdensities_comp.pdf}
%   \caption[]{\textit{Top:} SN energy injection rate over time. \textit{Center:} Hydrogen number density over time. \textit{Bottom:} Dark matter density in $40~\rm{pc}$, $100~\rm{pc}$ and $150~\rm{pc}$ over time. In the lower two panels the solid, dashed and dotted lines are measured within $40$pc, $100$pc and $150$pc, respectively. Both the gas density and dark matter react to the injection of SN energy, as features match between the two. However, it is not sufficient to form a core. }
%   \label{fig:DM_100pc}
% \end{figure*}
%%%%%%%%%%%%%%%%%%%%%%%%%%%%%%%%%%%%%%%%%%%%%%%%%%%%%%%%%%%%%%%%%%%%%%%%%%%

%%%%%%%%%%%%%%%%%%%%%%%%%%%%%%%%%%%%%%%%%%%%%%%%%%%%%%%%%%%%%%%%%%%%%%%%%%%
\subsection{Dark matter profile}
\label{sec:dmprofile}
%%%%%%%%%%%%%%%%%%%%%%%%%%%%%%%%%%%%%%%%%%%%%%%%%%%%%%%%%%%%%%%%%%%%%%%%%%%
Dark matter profiles of dwarf galaxies are of wide concern to the astrophysics community \citep[e.g.][]{Bullock2017}. The apparent presence of cored profiles in observations \citep{FloresPrimack1994, Moore1994, McGaugh2001, Read2017} has been taken to pose a challenge to the $\Lambda$CDM paradigm since dark matter-only simulations predict NFW \citep{Navarro1993} profiles with a central cusp. Various simulation projects have since shown how dark matter cores may be formed in response to baryonic oscillations in the centers of halos caused by repeated cycles of accretion and outflow \citep[e.g.][]{PontzenGovernato2012, PontzenGovernato2014, Teyssier2013, DiCintio2014, Onorbe2015, Tollet2016, Dutton2016, Orkney2021}.  

In Fig.~\ref{fig:dmprofile} we compare the dark matter density profiles with the corresponding ones in a dark matter-only (DMO) simulation. The DMO lines are re-scaled to account for the cosmic baryon fraction. The hatched region extends to $r_{2000}$, the radius that encloses 2000 DM particles. Following \cite{Power2003}, we assume the flattening of the profile within $r_{2000}$ is numerical. No distinct core is visible in any of the four simulations. To verify this, we fit the slope of the profiles in two radial ranges and present the fitting results in Tab.~\ref{tab:dmfit}. The first range is the commonly used range between $1\%$ and $2\%$ of the virial radius, which we also show in the plot between the two dotted grey lines. The second range is between $r_{2000}$ and $3\,r_{2000}$, which is shown in the plot between the end of the hatched region and the vertical solid grey line. The uncertainty of the profile is estimated for each logarithmic bin in radius assuming Poisson statistics, since the dark matter density is sampled by discrete massive particles. This uncertainty is passed on to the least-squares fit to obtain the errors for each slope shown in the table.
In the last column, we also present the stellar-to-halo mass ratio, for comparison. 

%The only indication of a slight, non-significant core is in the $10^6$ simulation when measured within $3r_{2000}$. None of the slopes in the $1-2\%$ range deviate from the DMO simulation within their respective errors. 
The $\alpha_{1-2\%}$ values of the simulations with baryons are consistently below the DMO run. However, these differences are slight and in most cases not larger than the 1-$\sigma$ uncertainties. This small effect can be interpreted as a tendency towards halo expansion at these radii ($300 \lesssim R/\mathrm{pc}\lesssim 600$). The $\alpha_{2000}$ values, on the other hand, probe the slope significantly closer to the center. Interestingly, these values are larger than the DMO value in all cases except for the ``$10^6$'' run. This is consistent with a slight halo contraction ($35 \lesssim R/\mathrm{pc}\lesssim 100$). The ``$10^6$'' run yields a lower slope but with a difference that is not more than the 1-$\sigma$ uncertainty. The differences between the profile shapes for the four models are small but indicative of the fact that the dark matter will react even to small differences in the stellar distribution. In all four models, the stars dominate the central potential out to $30-50\,$pc and contribute significantly to the mass out to $100-150\,$pc (cf. Fig.~\ref{fig:rotcurve}).  ``$10^6$'' has the shallowest stellar profile and the shallowest dark matter profile. The same can be said in reverse for ``$10^7$'', which has the steepest profile (regarding $\alpha_{2000}$). We conclude that there is no indication of core formation in these simulations, since the slopes do not suggest a constant density in the centre. Instead, the DM appears to merely follow slight changes in the stellar distribution, which in turn is sensitive to the assembly history of the central stellar spheroid.

According to previous simulation work \citep{Penarrubia2012, DiCintio2014, Tollet2016} the logarithmic stellar-to-halo mass ratio is a strong indicator of the central DM density slope. Our simulations display values between $-3.1$ and $-3.5$ for this ratio. According to \cite{DiCintio2014}, this range would suggest $\alpha_{1-2\%}$ values of $-0.6$ to $-1.2$, giving rise to distinctly shallower profiles than we see in our work. In contrast, the FIRE simulations exhibit steeper slopes in this same range ($\alpha_{1-2\%}$ values of $-1.1$ to $-1.5$, \citealt{Chan2015}), in agreement with our results. We now turn to the central DM and gas densities to gain further insight on this issue. 

In Fig.~\ref{fig:DM_100pc} we examine three quantities: in the top panels the SN energy injection rate, $\zeta_\mathrm{SN}$, in the middle panels the central gas density and in the lower panels the central DM density. For the two lower rows, we measure within $40\,$pc, $100\,$pc and $150\,$pc, shown in solid, dashed and dotted lines, respectively. In the lower panels we also show the DM density of the dark matter-only run for comparison (grey lines).
%We see that the SN rate clearly follows the SFR, as is to be expected. 
%
The DM density of the hydrodynamic run remains within a factor of $2-3$ of the DMO run at all times. This said, there are two deviations we would like to discuss. For one, in ``$10^7$'', after SF has ceased around $t\approx2\,$Gyr, the DM density remains increased. As this is not the case in the other models, it must be related to the PopIII implementation. As we can see from the right central panel, the gas density during the onset of SF in this model is significantly higher than in the other models. This in turn leads to a more pronounced collapse and a higher central stellar density, which draws more DM to the center, making the profile cuspier and the central DM density higher than in the DMO run (as we have similarly seen above).  

The second point we would like to mention here is the slow but steady reduction of DM density across $\sim10\,$Gyr in all four models. The reduction is strongest in ``$10^7$'', where the density is decreased by $3.5\times10^8\Msun\,\mathrm{kpc}^{-3}$ between $z=4$ and  $z=0$. In the other three models the reduction is  $\lesssim2.4\times10^8\Msun\,\mathrm{kpc}^{-3}$. Since no SF occurs within the halo between $t=2 - 10\,$Gyr, SF and outflows cannot be the cause. Instead, we conjecture this is due to minor mergers. \cite{Naab2009} and \cite{Orkney2021} have shown that this is a viable DM heating mechanism. As the reduction differs between the models and there is little gas in the halos for most of their evolution, the culprit must again be the stellar density. We clearly see that the ``\tkms'' model has a flatter evolution than its analogue, ``$10^6$''. As we will see in the next section, this model has significantly less infalling substructure. Fewer minor mergers thus lead to a smaller reduction in the central density. We conclude that while there are no detectable cores at $z=0$, the halos do slightly expand in response to the continual accretion of dense stellar substructure.
%
%Both the gas and DM density react to the SN injection by decreasing temporarily. However, they repeatedly recover their original value. ``$10^7$'' displays a lasting difference to the DMO run by having a slightly higher central DM density. The gas density is always highest leading up to a SF episode and then deceases rapidly when the SN rate increases. 

As we can see by inspecting the $y$-axis values in the central and lower panels, the gas never dominates the central potential. When it is highest it just about matches the DM density. Previous work \citep[e.g.][]{Fitts2017, Bose2019} has shown that the gas content within the center must be gravitationally dominant over the DM for a core to form. {{Our results are consistent with}} this requirement, since our simulations do not show cores and do not exhibit gas-dominated regions in the center.
\cite{Onorbe2015} and \cite{Fitts2017} discuss stronger core formation in the presence of late time SF (at fixed total stellar mass). However we cannot confirm this, as both ``homogeneous'' and ``$10^7$'' exhibit late bursts of SF that do not affect the central dark matter density. 

Additionally, we mention the results \citep{Governato2010, Dutton2019, Dutton2020} that show a dependence of core formation on the density threshold for SF. The authors argue that higher SF thresholds lead to larger (more spatially extended) cores. In simulations with a threshold of $0.01\mathrm{cm}^{-3}$ no cores form, while thresholds of $100-1000\,\mathrm{cm}^{-3}$ lead to large cores. We cannot confirm this result at face value as our threshold is $1000\,\mathrm{cm}^{-3}$ and our simulations do not create cores. Analyzing the central panels of Fig.~\ref{fig:DM_100pc} again, we can see that the central gas densities grow in the second half of the elapsed cosmic time. However, a central density of $\sim10^7 \Msun\mathrm{kpc}^{-3}$ is already sufficient to reignite SF. To dominate over the DM density, SF would have to be deferred until the gas density increases to $\gtrsim10^9 \Msun\,\mathrm{kpc}^{-3}$. This would require a drastically higher SF threshold of $\gtrsim10^5\mathrm{cm}^{-3}$.

We note two important caveats here. Firstly, our dwarf is a very low mass galaxy for which the addition of other forms of feedback may reduce the stellar mass and could, thus, affect the results presented here. Secondly, we are only considering a single galaxy simulated with some parameter variations. As such, we do not attempt to make any sort of statistical claim for the full population of dwarfs. A more detailed investigation is postponed to future work when more simulations will be available, while the present analysis only serves to outline the general properties of our initial \lyra simulations.

%%%%%%%%%%%%%%%%%%%%%%%%%%%%%%%%%%%%%%%%%%%%%%%%%%%%%%%%%%%%%%%%%%%%%%%%%%%
\subsection{PopIII enrichment and stellar substructure}
\label{sec:popiii}
\begin{figure}
  \includegraphics[width=\linewidth]{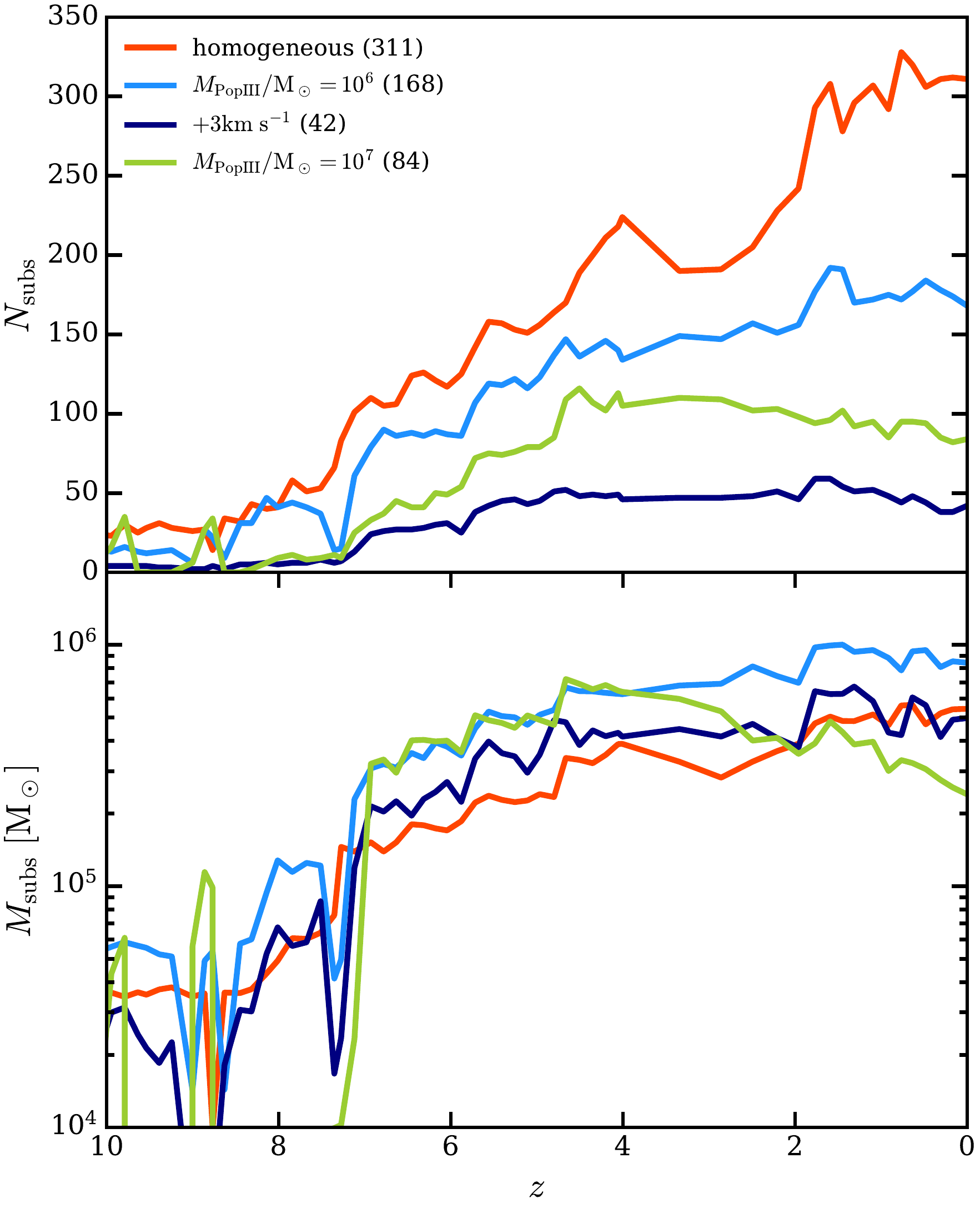}
  \caption[]{Number (\textit{upper panel}) and total mass (\textit{lower panel}) of luminous bound substructure within $R_{200}$ as a function of redshift. These are gravitationally bound objects found with the substructure finder {\small Subfind}. We used the following selection criteria: the number of particles must be 32 or larger and needs to include at least one star. 
  %On the right we added a mass cut of $>10^4\Msun$ and only counted structure within $0.2\Rvir$. These choices are intended to mimic the approximate detection limit of the Solo survey. 
  The number in parentheses in the legend is the value at $z=0$. The number of substructures increases rapidly after $z=7.5$ when reionization has set in ($z\approx8$) since SF is quenched and outflows have ceased to prevent accretion. The different models show very different predictions at late times.}
  \label{fig:nsubs}
\end{figure}
%%%%%%%%%%%%%%%%%%%%%%%%%%%%%%%%%%%%%%%%%%%%%%%%%%%%%%%%%%%%%%%%%%%%%%%%%%%
Finally, we highlight an exciting prediction concerning dwarf substructure that results from the differences between the four models. In the following, we use the term luminous substructure to denote gravitationally bound stellar objects of any mass, remaining agnostic about the presence or absence of a dark matter component.

In Fig.~\ref{fig:nsubs} we show the redshift evolution of the number {{and total mass}} of luminous, bound substructures within the virial radius. The substructures are found using {\small Subfind}. To include a {\small Subfind} object, we require that it is composed of 32 particles or more, at least one of which must be a star particle. 
%We also verify that it is spatially resolved by requiring a half mass radius greater than the stellar softening ($>4$pc). These criteria are intended to detect visible substructure. 
During the SF phase {{($10>z>7$)}}, the number of such objects is small. Then at $z\approx7$ the number increases rapidly in all four models. This can be easily attributed to the onset of reionization. As reionization quenches SF, the SNe that drive outflows soon cease, too. Once this happens accretion once again becomes the dominant motion. The majority of substructure is accreted after reionization and accumulates within the halo. This is mirrored in an ex-situ fraction of $>95\%$ in all models {{(fraction of stars formed outside the main halo)}}.

The different models show a very different evolution for their substructures. The ``homogeneous'' model hosts many more luminous subhalos than the other three models. {{However, as we can see in the lower panel, the total mass in substructure is not increased. This means most of these are low mass objects.}} The increased availability of enriched (and cooling) gas allows for many small clumps to condense and form stars. But the differences become more pronounced at later times around $z\approx4-0$. In the ``homogeneous'' model, we see the accretion of stellar substructure continues after the other models have slowed their growth. This can be attributed to the widespread presence of small luminous halos in the vicinity of the main halo that are not as prevalent in the other models (see also Fig.~\ref{fig:firstsubs}). 

% \cite{Hislop2021} lack of correct disruption of star clusters, small systems too dense
% \cite{Lahen2020, Li2021} isolated galaxy mergers with cluster formation.

With increasing $M_\mathrm{PopIII}$ the number of luminous objects available for accretion declines. This is well reflected in the reduction at $z=0$ from $311$ (homogeneous) to $168$ ($10^6$) and finally to $84$ ($10^7$) over-densities within $R_{200}$. This includes small objects with total masses down to $<100\Msun$ {{and is not mirrored in the in total mass of substructure}}. The lowest number is attained by the ``\tkms'' run with $42$. This shows that many objects found in the other runs {{are extremely low mass}}. If the stars have marginally higher thermal velocities (as in the ``\tkms'' run), these objects {{do not remain self-gravitating but instead disipate and merge with the central object. Indeed, the ``$10^7$'' run shows a reduction in number and mass of substructures after $z\approx4$. The high central stellar density gradient in this simulation contributes to  tidally disrupting orbiting over-densities (cf. Fig.~\ref{fig:rotcurve})}}. We also note that there are many objects well below the $M_\mathrm{PopIII}$ threshold. As discussed in Sec.~\ref{sec:model} many of them are formed ex-situ with the aid of metal enrichment from neighboring halos. They are later accreted by the main halo to join the many luminous objects we see at $z=0$.

This preliminary result opens a compelling prospect of constraining the halo masses of PopIII stars with observations of high redshift dwarf galaxies and their substructure, which may in fact become possible in the next few years. According to \cite{Patej2015}, the high redshift progenitors of some of the brightest LG dwarf galaxies may be able to be detected in deep James Webb Space Telescope (JWST) observations if their stars form before $z\approx7-6$. According to our work, $>90\%$ of the stars should fulfill this requirement. Also, \cite{Schauer2020} make predictions for the observability of the first luminous halos.

We ask the reader to bear in mind, however, that thus far this is a purely theoretical prediction. Making a quantitative, redshift-dependent statement about the number of detectable structures requires various further calculations. Firstly, we need to distinguish between star clusters and small satellites that have a dark matter component. Next, we require a dedicated analysis to confirm that the star clusters actually follow the cluster mass function across time, namely that we correctly resolve their formation, survival and destruction via two-body relaxation \citep{Spitzer1987}. Lastly, detailed mock surface brightness estimates for each subhalo are necessary to understand how many of these objects may feasibly be observed. These calculations will be the subject of a future study.
%==========================================================================%
% \section{Discussion}
\label{sec:discussion}
%==========================================================================%
% \input{discussion}
%==========================================================================%
\section{Discussion \& Conclusions}
\label{sec:conclusion}
%==========================================================================%
%

In this paper we present the results of the first cosmological \lyra simulations. These are of an isolated dwarf galaxy with a halo mass of $2\times10^9\Msun$, similar in mass to Local Group dwarf galaxies. The dwarf produces $\sim10^6\Msun$ in stars, of which $>90\%$ are formed before the onset of reionization at $z=8-7$.

We test three different models for introducing the first metals at high redshift. In the first scenario, we simply begin the simulations setting all gas metallicity to the metallicity floor of [Fe/H]~$= -4$. The other three models can be considered sub-grid models for Population III enrichment. We execute the halo finder on-the-fly and increase the metallicity from zero to the metallicity floor inside each halo that crosses a mass threshold, $M_\mathrm{PopIII}$. We test this model for two values of $M_\mathrm{PopIII}$, $10^6\Msun$ and $10^7\Msun$. Lastly, we consider a model where we compensate for unresolved stellar velocities by adding a small additional kick of \tkms. 

As regards to our results, we can confirm that the onset of reionization in the form of a redshift-dependent UV background is able to quench SF in low-mass dwarfs. There is a very clear connection between the cessation of SF and the increase in the UVB heating. Dense cores of cold, self-shielded gas are able to survive for a few megayears after reionization, providing a small number of new stars. However, these clouds are soon heated and expelled by SNe. Through the strong expulsion of gas caused by the SNe, a very low baryon fraction of $\approx0.01$ is established before the end of reionization. These SN-driven outflows allow gas to escape the halo's shallow potential well, enriching the inter-galactic medium well beyond the virial radius of our dwarf. Thus, interestingly, PopII stars may be the first stars to form in certain halos that were enriched by SNe from neighboring halos.

We do not find any statistically significant flattening of the DM density profiles. While there is some halo expansion due to accreted substructure, the central region cannot be considered to have a constant density. This result is somewhat in tension with previous predictions, since the stellar-to-halo mass ratio is fairly high (for a dwarf) and our SF density threshold is also high. However, we do refrain from making far reaching claims, since we have only analysed a single object, albeit with different model parameters.

We find two avenues for re-igniting SF. Mergers can restart SF for short times, but the galaxy is prone to become quenched again due to the new generation of SNe. Moreover, slow accretion of gas over many gigayears can reignite sustained SF at extremely low rates of $10^{-6}-10^{-5}\Msun \mathrm{yr}^{-1}$. This is contingent upon the central potential being sufficiently deep. In our model this depth is generated by a central stellar over-density, tentatively a nuclear star cluster.

The total halo mass and the stellar mass are not significantly affected by these model differences. However, while the total stellar mass is robust, the star formation history is slightly varied. The smaller the first halos, the earlier star formation commences. Additionally, there are two small bursts of star formation after reionization, at $z\approx 4$ and $0.1$. In the ``$10^6$'' and ``\tkms'' models, these do not appear. This leads us to our first major conclusion about the differences between the PopIII models. Higher halo masses of PopIII hosts collapse monolithically, creating higher central stellar densities than in smaller halos where the collapse is more distributed. Thus, the models that do not re-ignite SF do not do so because they have lower central densities. Rather, their assembly history is comprised of smaller, lower density halos. This in turn is insufficient to accrete and retain sufficient gas to re-ignite SF.

PopIII enrichment also has a non-negligible effect on the assembly history of the dwarf in a second way. It affects the number of accreted substructures that host stars. Thus, we can tentatively propose that the number of luminous substructures in LG dwarfs may be able provide constraints on the masses of halos that form the first stars. This has the additional effect that the ex-situ fraction of our $\approx10^9\Msun$ dwarf galaxy is $>95\%$ since it forms hierarchically from smaller halos hosting stars.

Another difference between the enrichment models is the metallicity distribution function which is either approximately uni-modal in the ``homogeneous'' model or bi-modal in the ``$10^6$'' model. The bi-modality is caused by the fact that each $10^6\Msun$ halo forms a short burst of SF when it is initially enriched. But due to the small halo sizes, the bursts are short and are not able to feed multiple generations of stars that would become continually more enriched, thus moving the distribution towards uni-modality.

It is also interesting to compare our simulations to the EDGE \citep{Agertz2020} halos presented in \cite{Orkney2021}. The authors maintain that their simulations display cores while also not reaching gas densities that dominate over the DM.
The three lowest mass halos in their work are very similar to our galaxy in various ways but show visible cores out to $30-300\,$pc. These simulations are run with the EDGE model \citep{Agertz2020} which also resolves the multi-phase ISM and individual SN blasts. The gas mass resolution is only slightly worse than ours, with $18\Msun$ and a spatial resolution of $\sim3\,$pc. Likewise, the stellar mass formed in their three halos is between $10^5 - 10^6\Msun$, which is similar or only slightly lower than our final stellar mass. Thus, we cannot readily account for the disagreements seen in the DM profiles (if indeed disagreements remain when a larger sample is considered). However, SF and feedback implementations are likely cuplrits. We stress that this work only presents a single system. Thus, we refrain from making any statistical claims about the full dwarf population and defer a more detailed investigation to future work when more simulations are available. It is clear, however, that understanding the early enrichment is vital to make robust predictions on this mass scale of dwarf galaxies.

%
%==========================================================================%
\section*{Acknowledgments}
We thank the anonymous referee for aiding us in making this manuscript
more clear.
TN acknowledges
support from the Deutsche Forschungsgemeinschaft (DFG, German
Research Foundation) under Germany’s Excellence Strategy –
EXC-2094 – 390783311 from the DFG Cluster of Excellence
‘ORIGINS’.
This research was carried out on the High Performance Computing resources of the {\small FREYA} and {\small COBRA} clusters at the Max Planck Computing and Data Facility (MPCDF) in Garching operated by the Max Planck Society (MPG).
%==========================================================================%
\section*{Data availability}
The data underlying this article were accessed from the High Performance Computing resources of the Max Planck Computing and Data Facility (MPCDF, \url{https://www.mpcdf.mpg.de}). The derived data generated in this research will be shared on reasonable request to the corresponding author.
%==========================================================================%
%
%\renewcommand{\refname}{REFERENCES}
\bibliographystyle{mnras}
\bibliography{bib}

%
%\appendix
%
\bsp
\label{lastpage}
\balance
\end{document}